\documentclass[english]{article}
\usepackage{jcappub}
\usepackage{lmodern}

\usepackage{graphicx}
\usepackage{esint}
\usepackage{babel}

\makeatletter

\begin{document}
\title{Bayesian inference on the sphere beyond statistical isotropy}
\author[a]{Santanu Das}
\author[b]{Benjamin D. Wandelt}
\author[a]{Tarun Souradeep}

\emailAdd{santanud@iucaa.ernet.in}
\emailAdd{tarun@iucaa.ernet.in}
\emailAdd{wandelt@iap.fr}

\affiliation[a]{IUCAA, P. O. Bag 4, Ganeshkhind, Pune 411007, India}
\affiliation[b]{IAP, Lagrange Institute, Sorbonne University, Paris}


\abstract{
We present a general method for Bayesian inference of  the underlying covariance structure of random fields on a sphere. 
We employ the Bipolar Spherical Harmonic (BipoSH) representation of general covariance structure  on the sphere. We illustrate 
the  efficacy of the method as a principled approach to assess violation of statistical isotropy (SI) in the sky maps of Cosmic Microwave 
Background (CMB) fluctuations. SI violation in observed  CMB maps  arise due to known physical effects  such as Doppler boost and weak lensing; yet unknown theoretical possibilities like cosmic topology and subtle violations of the cosmological principle, as well as, expected observational artefacts of  scanning the sky with a non-circular beam, masking, foreground residuals, anisotropic noise, etc.  We explicitly demonstrate 
the recovery of  the input SI violation signals with their full statistics in  simulated CMB maps. Our formalism easily adapts to exploring parametric physical models with non-SI covariance, as we illustrate for the inference of the parameters of a Doppler boosted sky map. Our approach promises to provide a robust quantitative evaluation of the evidence for SI violation related anomalies  in the CMB sky by estimating the BipoSH spectra along with their complete posterior. 
}

\maketitle

\section{Introduction}


Many astronomical measurements on sky, as well as, data from other other parts of science, such as, geophysical  and 
environmental modelling on the globe,  deal with analysis of random fields on a sphere.  Advancement in  observational data has opened up the possibility of  posing deeper inference problems with sophisticated analysis tools.
In this paper we present a novel way to implement Bayesian inference of the underlying covariance structure. 
We develop our analysis strongly motivated by the analysis of cosmic microwave background (CMB) sky maps to infer the statistical 
isotropy (SI) of our observed universe. However, the  methodology is  applicable to Bayesian inference for other cosmological studies, and  more widely to  any study involving a scalar random field on the sphere.

For a statistically isotropic Gaussian random field, the two-point correlation function is rotationally invariant 
and hence the covariance matrix of the corresponding  random spherical harmonic coefficients, i.e., 
$\left\langle a_{lm}^{*}a_{l'm'}\right\rangle $ is diagonal and independent of the azimuthal multipole index $m$. 
Therefore, the angular power spectrum ($C_{l}=\left\langle a_{lm}^{*}a_{lm}\right\rangle $) can alone provide
a complete statistical representation of the field on the sphere. However, in presence of SI violation, the covariance matrix 
$\left\langle a_{lm}^{*}a_{l'm'}\right\rangle $ can depend on $m$ and the off-diagonal components can be nonzero. 
This calls for a representation of the random field that 
generalizes the angular power spectrum. Bipolar Spherical harmonic representation (BipoSH) was  
first proposed by Hajian and Souradeep~\cite{Hajian2003} to analyze SI violation signals on CMB. Any SI violation 
of a random field on a sphere can be completely represented in BipoSH spectra
($\tilde{A}_{ll'}^{LM}$)~\cite{Das2013,Das2014,Joshi2012,Souradeep2001}. The $L=0$ component of the BipoSH spectra, $\tilde{A}_{ll}^{00}$, is the angular power spectrum, $C_{l}$.

The angular power spectrum $(C_{l})$ can be estimated from the observed 
sky map using Bayesian statistics. Earlier Maximum likelihood methods~\cite{Bond1998,Efstathiou2003,Gorski1994,Oh1999,Ben2004} were 
used for optimal estimation of the CMB power spectrum. A straight forward minimisation of the likelihood is computationally  prohibitive 
as it require $O(N_{pix}^{3})$ computational complexity, where $N_{pix}$ is the number of pixels present in the map. 
Last maximum likelihood methods on CMB dataset were used in some parts of the Boomerang and Maxima datasets. Since $N_{pix}$
increased in the modern experiments, maximum likelihood methods are not a viable option and pseudo-$C_l$ methods~\cite{Ben2001,Hivon2002,Brown2004} are used since then.  
We can explore the full posterior distribution with Monte Carlo sampling. However, this approach relies of the availability of 
an efficient sampling method. At low multipoles where the posterior of the $C_l$ are highly non-Gaussian due to the small number of degrees of 
freedom, Gibbs sampling~\cite{Ben2004} is now standard. Other sampling methods, used for sampling the posterior of the CMB power spectrum, involves methods like Hamiltonian Monte Carlo~\cite{Taylor2008} etc. 

For a SI violated Gaussian random field on a sphere, calculating the only the angular power spectrum  by
minimising the likelihood with respect to $C_{l}$ is not sufficient.
Although, when the SI violation is small in comparison
to the $C_{l}$, the posterior distribution of the $C_{l}$ may not
alter significantly, it is still important to minimise the likelihood with
respect to the off-diagonal components of the covariance matrix  (represented in terms of BipoSH spectra, $\tilde{A}_{ll'}^{LM}$), 
which not only guarantees complete posterior statistics of $C_{l}$ but also provides a complete posterior statistics of 
BipoSH spectra. We present a novel method for estimating 
the posterior distribution of BipoSH spectra by sampling the full covariance matrix.
The Bayesian method for finding the posterior distribution of  BipoSH spectra allows a complete and reliable statistical inference of 
the presence of SI violation over the sphere. For sampling the posterior, we choose the Hamiltonian Monte Carlo~\cite{Hajian2007,Duane1987,Taylor2008}, 
a Monte Carlo method that uses the classical Hamiltonian mechanics for calculating the posterior 
distribution and is capable of sampling the posterior distribution faster than other conventional 
MCMC methods in such high dimensional problems. 


There has been a phenomenal improvement in CMB observations in the past few years. 
After recent data release of WMAP-9 and Planck, the observations in CMB have achieved the level of reliability where  every single significant departure from the standard model cosmology may present a challenge to our understanding of the universe. 
The standard homogeneous and isotropic cosmological model dictates that the temperature fluctuation in CMB sky respect 
Statistical isotropy (SI). However, recent data from  Planck hint at  some putative signals of SI violation 
in the CMB sky~\cite{wmap7-anomalies,wmap9-anomalies,planck_isotropy}. There are several theoretical models proposed 
in literature that  lead to SI violation in the CMB such as the non-trivial cosmic topology \cite{Ellis1971,Bond1998,Bond2000,Bond2000a}, anisotropic cosmology,
etc.~\cite{Ackerman2007,Pullen2007}. Weak lensing of CMB and Doppler boost are among the known 
effects that lead to SI violation in the observed CMB~\cite{AR-MA-TS-lens-biposh,Suvodip2014}.  Artefacts  of  observational reality such as anisotropic noise,  foreground residuals from inadequate  cleaning, effect of non-circular beam response function, etc.~\cite{Das2013,Das2014,Joshi2012} may also yield similar SI violation. Therefore, it is necessary to extract the SI violation signal  from  observational data accounting for any false signal of observational systematics, and then  carefully statistically assess its significance vis a vis peculiarity of a  particular realisation of  SI 
covariance. We apply our method of the Bayesian inference of random fields on sphere to analyse simulated  CMB maps and 
recover  input isotropy violation signals  in the simulated CMB sky. Our method  is shown to successfully recover the input signal properly up to high multipoles. Another important fact is that our Bayesian approach produces inferences that are data-dependent; i.e. they allow testing the relative probability of different models given the realized data set. This is in contrast to frequentist approaches that compare the data set to an ensemble of alternate data sets generated from the null hypothesis.

The paper is organised as follows. In section~\ref{sec_biposh}, we provide a brief review of the BipoSH formalism and  present the  
likelihood for the full covariance matrix.  Section~\ref{hmc} discusses the details of the equation of motion for
the HMC method. In  section~\ref{compute}, we provide  details of our handling of the  computational issues. The analysis of simulated CMB sky maps and results are included in the section~\ref{results} of the paper. Here, first we do our analysis on statistically isotropic CMB maps and show that recovered BipoSH spectra are consistent with $0$. Then we show our analysis on SI violated sky maps  originated from WMAP beam and scan pattern, dipolar modulation and the Doppler boost signal originated from the motion of our galaxy respectively. In this section, we also develop and method to explicitly recover the Doppler boost signal by directly sampling the posterior. The final section~\ref{concl} is devoted to discussions and conclusions.

\section{A brief review of BipoSH representation}
\label{sec_biposh}


Consider  measurements  that are  a linear transform of a random signal and  an additive instrumental
noise, all  being fields defined on a sphere. We pose  the general problem of Bayesian inference 
of the underlying covariance structure of the signal field, given the knowledge of the noise covariance 
 and the linear transformation relating  the measured data to the  signal. 

With no loss of any generality of the problem, we specifically consider  the measurement of 
CMB temperature  fluctuations in the sky. The observed sky map is a convolution of the real sky temperature with the 
instrumental beam with an addition of  instrumental noise. Therefore, $\tilde{T}(\gamma)$  the actual temperature 
signal of the CMB sky along the direction $\gamma$ is  linearly related  to  the observed sky temperature, $\tilde{d}(\gamma)$,   as

\begin{equation}
\tilde{d}(\gamma)=\int B(\gamma,\gamma')\tilde{T}(\gamma')d\Omega_{\gamma'}+\tilde{N}(\gamma)\,,
\end{equation}

\noindent where $B(\gamma,\gamma')$ is the instrumental beam profile and $\tilde{N}(\gamma)$
is the instrumental noise. For a perfectly circular 
beam profile, $B(\gamma,\gamma')\equiv B(\gamma\cdot\gamma')$, assumed in this work,  it is easy 
to deconvolve the effect of the beam after inferring the power spectra. However, if the beam is not circular symmetric 
then the effect of the beam depends on the full scan pattern of the
experiment and its deconvolution may be 
non-trivial~\cite{Das2014,Joshi2012,Armitage2004,Keih2012,Keih2015}. 

For data defined on a sphere, it is convenient to work in the spherical harmonic
space. The CMB signal,  $\tilde{T}(\gamma)$, then can be expanded in
terms of spherical harmonics as 

\begin{equation}
\tilde{T}(\gamma)=\sum_{l=0}^{\infty}\sum_{m=-l}^{l}a_{lm}Y_{lm}(\gamma)\,,
\end{equation}

\noindent where $Y_{lm}(\gamma)$  are the spherical harmonic functions and
$a_{lm}$ are the coefficients in the spherical harmonic basis. Similarly,
the observed data, $\tilde{d}(\gamma)$, can also be expanded in spherical
harmonics with coefficients, $d_{lm}$. 

For a perfectly statistically isotropic sky, the two point correlation
function on sky can be expressed in terms of the angular power spectrum,
$C_{l}$,  alone as 

\begin{equation}
\left\langle a_{lm}a_{l'm'}^{*}\right\rangle =S_{lml'm'}=C_{l}\delta_{ll'}\delta_{mm'}\,.
\end{equation}

\noindent Here $\left\langle \,\ldots\,\right\rangle $ denotes the ensemble
average. However, when we allow for CMB to have SI violation, $C_{l}$ does
not  provide a full description of the covariance matrix.
A general covariance matrix can be expanded in the BipoSH representation
as 

\begin{equation}
\left\langle a_{lm}a_{l'm'}^{*}\right\rangle =\sum_{m,m'}\left(-1\right)^{m'}A_{ll'}^{LM}\mathcal{C}_{lml'-m'}^{LM}\,,
\label{eq:BipoSH}
\end{equation}

\noindent where $\mathcal{C}_{lml'-m'}^{LM}$ are the Clebsch Gordon coefficients
and $A_{ll'}^{LM}$ are the BipoSH spectra that provide a natural generalisation of 
the angular power spectrum. Since, $\mathcal{C}_{lml'-m'}^{LM}$
span the entire space of the Covariance matrix, given a set of BipoSH
spectra we can calculate the entire covariance matrix and vice
versa. It is more convenient for most non-SI phenomena to use the even-parity BipoSH spectra 
$\bar{A}_{ll'}^{LM}=\frac{\sqrt{2L+1}}{\sqrt{2l+1}\sqrt{2l'+1}}\frac{1}{\mathcal{C}_{l0l'0}^{L0}}A_{ll'}^{LM}$
 that  more  closely  generalise  $C_l$.   The biposh representation splits the covariance matrix into pieces that transform separately under the action of the group of rotation $SO(3)$. The scalar (rotationally invariant) term is the power spectrum, $C_l$.


The goal of this paper is to calculate the posterior distribution of $A_{ll'}^{LM}$
from the observed sky map, i.e.,  $P(A_{ll'}^{LM}|d_{lm})$ or $P(S_{lml'm'}|d_{lm})$.
 Rather than computing this pdf directly
we sample the joint probability distribution, $P(S_{lml'm'},a_{lm}|d_{lm})$,  
and then marginalise over $a_{lm}$. The joint distribution can be obtained directly
by using Bayes Theorem~\cite{Bond1998,Tegmark1996,Gorski1996,Gorski1994}

\begin{eqnarray}
P(S_{lml'm'},a_{lm}|d_{lm}) & = & P(d_{lm}|a_{lm})P(a_{lm}|S_{lml'm'})P(S_{lml'm'})\nonumber \\
 & = & \frac{1}{\sqrt{|N_{lml'm'}|}}\exp\left[-\frac{1}{2}\sum_{lml'm'}\left(d_{lm}^{*}-a_{lm}^{*}\right)^{T}N_{lml'm'}^{-1}\left(d_{l'm'}-a_{l'm'}\right)\right]\nonumber \\
 &  & \times\frac{1}{\sqrt{|S_{lml'm'}|}}\exp\left(-\frac{1}{2}\sum_{lml'm'}a_{lm}^{*T}S_{lml'm'}^{-1}a_{l'm'}\right)P(S_{lml'm'})
\end{eqnarray}

\noindent where $N_{lml'm'}^{-1}$ and $S_{lml'm'}^{-1}$ are the elements of the inverse matrix of 
$N_{lml'm'}$ and $S_{lml'm'}$ respectively.  $P(S_{lml'm'})$ is the prior on $S_{lml'm'}$. The choice of $P(S_{lml'm'})$
has been studied in literature for the case of SI skymaps~\cite{Ben2004}.
Here, for our analysis we use a flat prior on $S_{lml'm'}$, i.e.,  $P(S_{lml'm'})=1$. However, other choices of prior 
such as Jeffreys prior in particular can also be used for the alanysis.

\section{Hamiltonian Monte Carlo Sampling of BipoSH spectra}
\label{hmc}

Conventional Monte Carlo techniques, such as Gibbs sampling or Metropolis
Hastings,  draw samples from a given probability distribution $P(S_{lml'm'},a_{lm}|d_{lm})$
using a random walk. On the other hand,  Hamiltonian Monte Carlo
(HMC) technique based on the Classical Hamiltonian Mechanics relies on the fact
that  the density of a  group of particles with random momenta  placed in a potential
will trace the potential  given that all of them start from random momentum drawn from a normal 
distribution with mean $0$ and co-variance $M$~\cite{Hajian2007,Taylor2008,Neal2012,Hoffman2011,
Duane1987}, where $M$ is a positive definite matrix called the mass matrix and can be choosen independently. It is known that  HMC method can sample the distribution
more effectively even in very high dimensional space
in comparison to other conventional MCMC methods. 

We sample the distribution $P(S_{lml'm'},a_{lm}|d_{lm})$ using HMC
with the parameters $a_{lm}$, $A_{ll'}^{LM}$. In a Hamiltonian Monte
Carlo algorithm we need to define a conjugate momentum and a mass
corresponding to each of its parameters. We consider the conjugate
momentum to $a_{lm}$ and $A_{ll'}^{LM}$ to be $p_{a_{lm}}$ and $p_{A_{ll'}^{LM}}$
and a corresponding mass $m_{a_{lm}}$, $m_{A_{l_{1}l_{2}}^{LM}}$ respectively. 
The mass matrices are the positive definite quantity by their definition.  
The potential in the Hamiltonian is taken as $-\log P(S_{lml'm'},a_{lm}|d_{lm})$ which leads the HMC sampler 
to sample the posterior of   $ P(S_{lml'm'},a_{lm}|d_{lm})$~\cite{Neal2012,Duane1987}. Thus the Hamiltonian for the motion of this ensemble
of particles is 

\begin{equation}
H=\sum_{lm}\frac{p_{a_{lm}}^{2}}{2m_{a_{lm}}}+\sum_{LMll'}\frac{p_{A_{ll'}^{LM}}^{2}}{2m_{A_{ll'}^{LM}}}-\ln(P(S_{lml'm'},a_{lm}|d_{lm}))\,.
\end{equation}

\noindent Using Hamiltonian mechanics, the equations of motion for $a_{lm}$
can be written as

\begin{equation}
\dot{a}_{lm}=p_{a_{lm}}/m_{a_{lm}}\label{eq:almdot}
\end{equation}

\noindent and 

\begin{eqnarray}
\dot{p}_{a_{lm}} & = & -\frac{\partial H}{\partial a_{lm}}\,=\sum_{l'm'}N_{lml'm'}^{-1}\left(d_{l'm'}^{*}-a_{l'm'}^{*}\right)-\sum_{l'm'}S_{lml'm'}^{-1}a_{l'm'}^{*},\label{eq:pdot}
\end{eqnarray}

\noindent Similarly, the equations of motion for $A_{ll'}^{LM}$ will be 

\begin{equation}
\dot{A}_{ll'}^{LM}=p_{A_{ll'}^{LM}}/m_{A_{ll'}^{LM}}
\end{equation}

\noindent and 

\begin{equation}
\dot{p}_{A_{ll'}^{LM}}=-\frac{\partial H}{\partial A_{ll'}^{LM}}=-\frac{1}{2}\partial_{A_{ll'}^{LM}}\ln\left|S\right|+\partial_{A_{ll'}^{LM}}\left(\sum_{lml'm'}a_{lm}^{*}S^{-1}_{lml'm'}a_{l'm'}\right)\,. \label{eq:ALMll}
\end{equation}

\noindent Using Eq.(\ref{eq:BipoSH}) and the orthogonality properties of the
Clebsch Gordon coefficients~\cite{Verselowich} we  obtain 

\begin{equation}
\partial_{A_{ll'}^{LM}}\left(\sum_{lml'm'}a_{lm}^{*}S^{-1}_{lml'm'}a_{l'm'}\right)=\sum_{mm'}C_{lml'm'}^{LM}\left(S^{-1}a\right)_{lm}\left(S^{-1}a\right)_{l'm'}\label{eq:expterm}
\end{equation}

\noindent and

\begin{equation}
\partial_{A_{ll'}^{LM}}\ln\left|S\right|={\rm tr}\left(\frac{\partial S}{\partial A_{ll'}^{LM}}S^{-1}_{lml'm'}\right)=\sum_{mm'}C_{lml'm'}^{LM}S^{-1}_{lml'm'}\,.\label{eq:inverse}
\end{equation}

\noindent Here $\mbox{tr}(\,\ldots\,)$ represents the trace of the enclosed matrix.

HMC is performed in two steps. In the first step,  values of the
momentum variables are chosen from the Gaussian distribution of mean
$0$ and variance $m_{x}$, where $x\in(a_{lm},A_{ll'}^{LM})$. In
the next step a Metropolis update is performed from the state $(p_{a_{lm}},p_{A_{ll'}^{LM}},a_{lm},A_{ll'}^{LM})$
to a new state $(p_{a_{lm}}^{*},p_{A_{ll'}^{LM}}^{*},a_{lm}^{*},A_{ll'}^{LM*})$
by integrating the equations of motion through a time interval
of fixed size,   $\Delta t$. The Hamiltonian is computed in this new state and
the state is accepted with probability $\mbox{min}(1,\exp(-\Delta H))$,
where $\Delta H$ is the change in the Hamiltonian between these two
states. If the new state is accepted then similar operation is performed
considering $(a_{lm}^{*},A_{ll'}^{LM*})$ as the new position variable,
otherwise the position is not updated from $(a_{lm},A_{ll'}^{LM})$.
Choice of $m_{a_{lm}}$ and $m_{A_{ll'}^{LM}}$ decides the stability of
the integration process. HMC algorithm in general uses the Leapfrog
integration algorithm due to its time reversal symmetry and almost symplectic
nature. For $C_{l}$ inference,  the Leapfrog integration works well.
However, when the the covariance matrix $S_{lml'm'}$ is non-diagonal, we find that the 
Leapfrog integration method diverges and needs very small step size
for stable integration. Instead a fourth order Forest and Ruth integrator~\cite{FOREST1990}, which is a symplectic integrator 
that involves three Leapfrog steps, works better then a simple Leapfrog. Therefore, we use fourth order Forest and Ruth 
integrator to integrate the dynamical equations..

\begin{figure}[h]
\centering
\includegraphics[width=0.45\textwidth,trim = 0 245 0 255, clip]{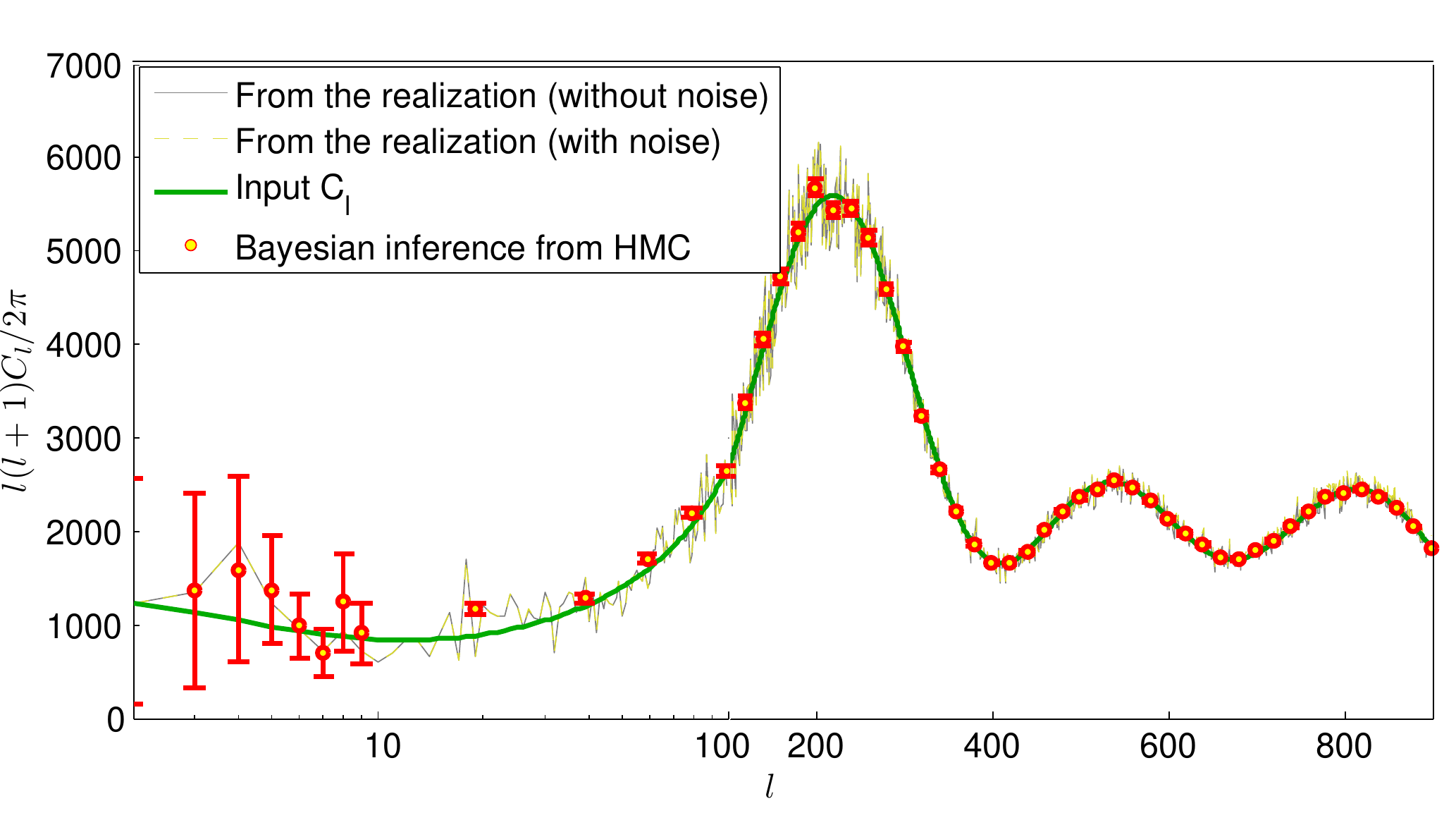}          
\includegraphics[width=0.45\textwidth,trim = 0 245 0 255, clip]{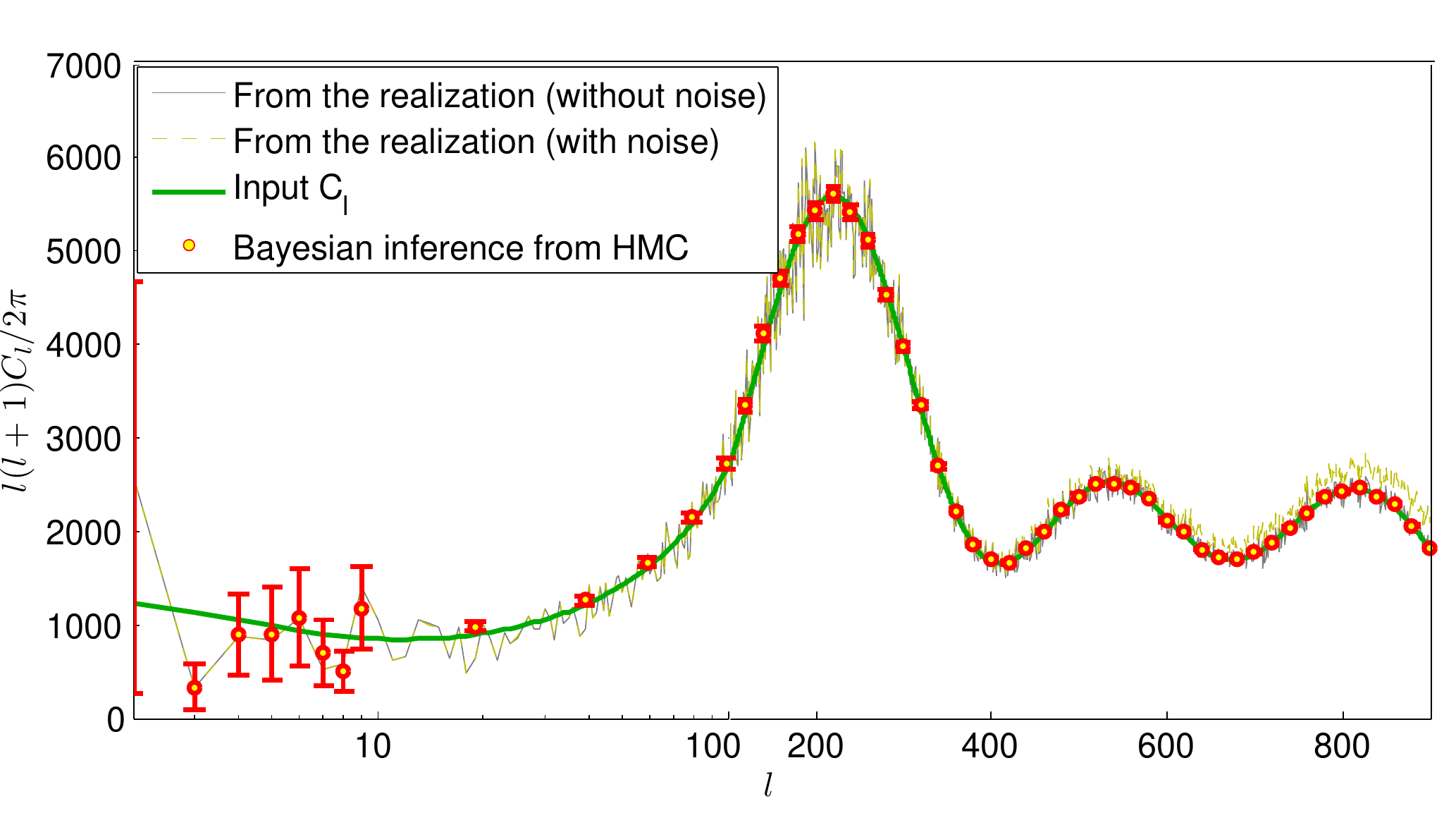}

\includegraphics[width=0.45\textwidth,trim = 0 245 15 255, clip]{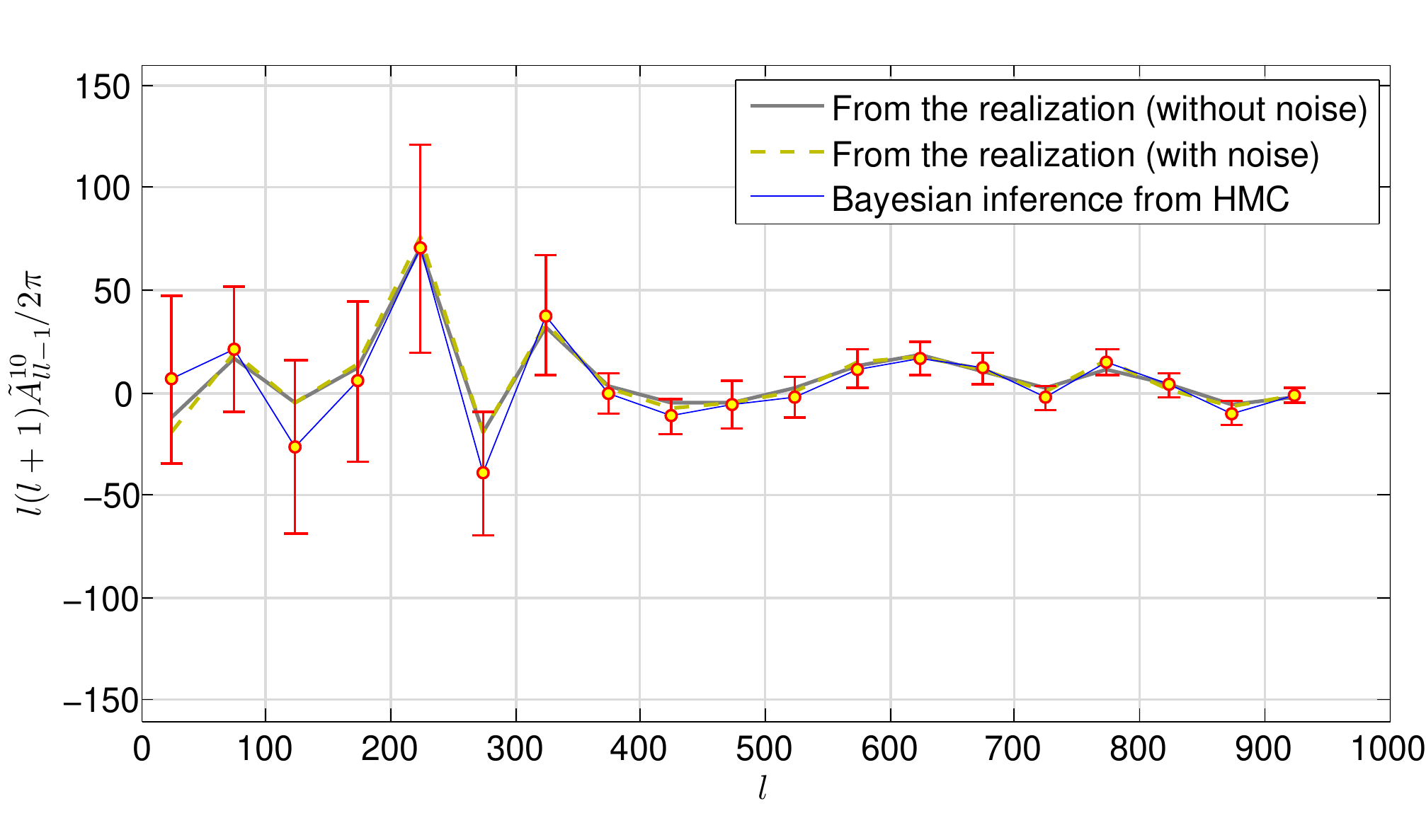}          
\includegraphics[width=0.45\textwidth,trim = 0 245 15 255, clip]{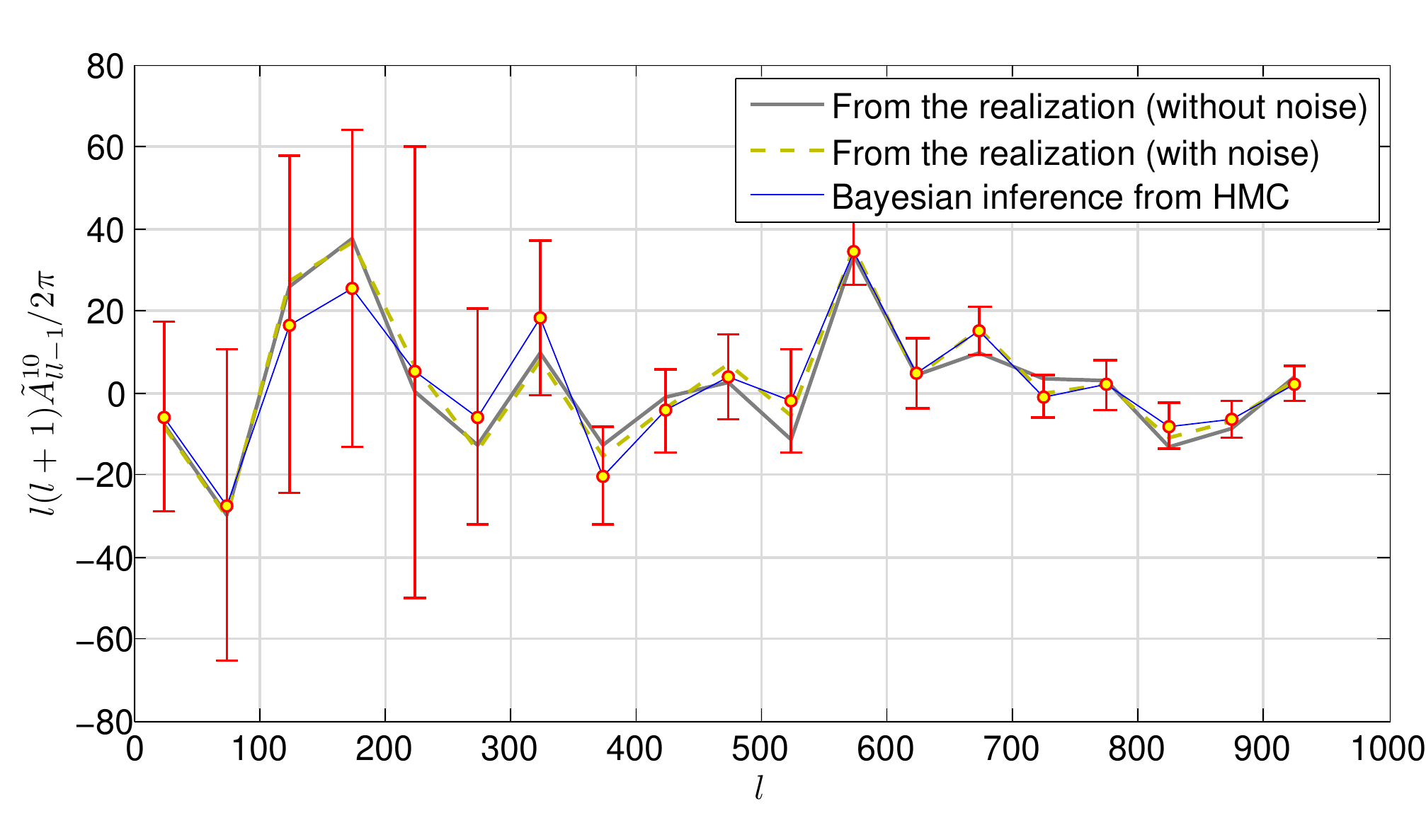}

\includegraphics[width=0.45\textwidth,trim = 0 245 15 255, clip]{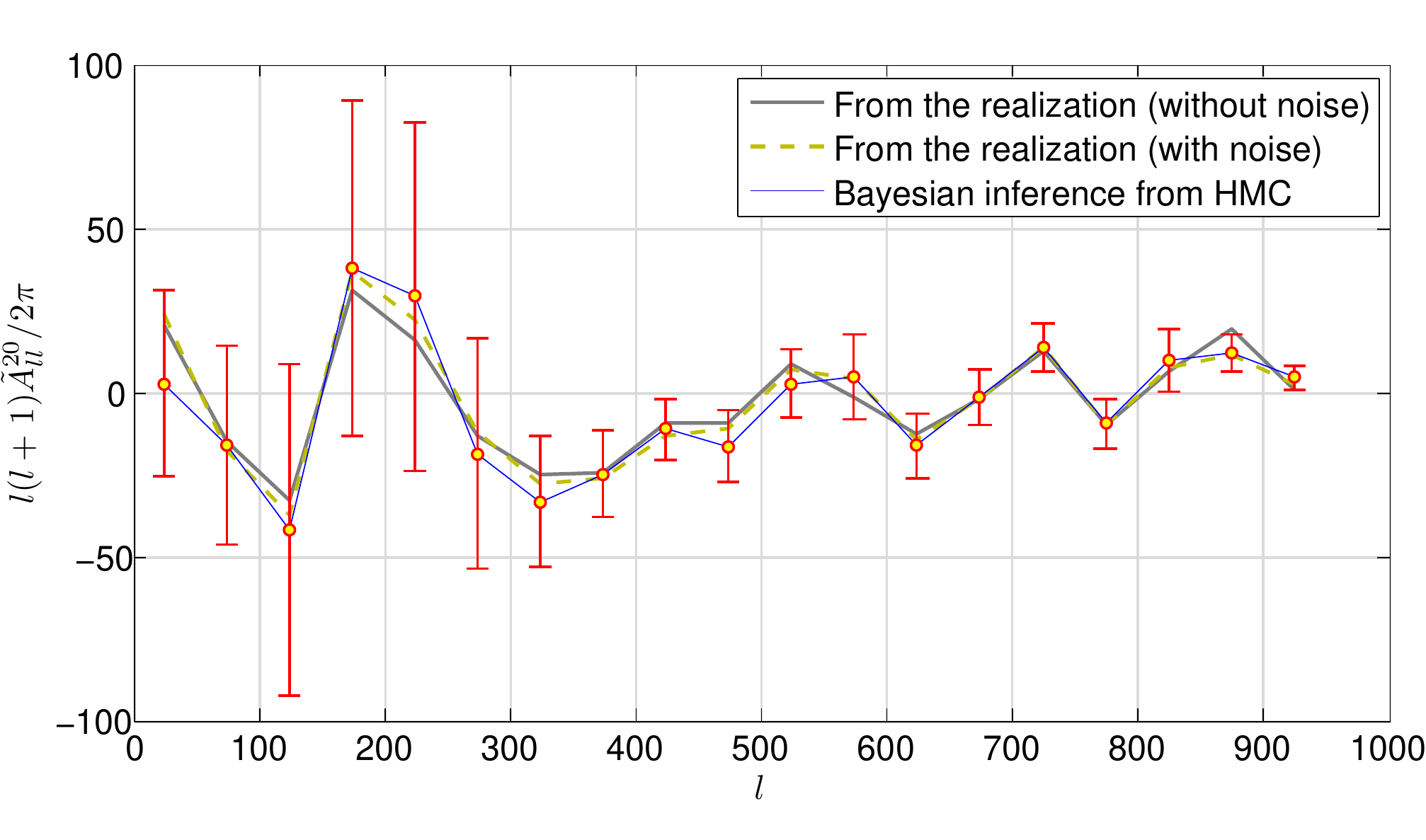}          
\includegraphics[width=0.45\textwidth,trim = 0 245 15 255, clip]{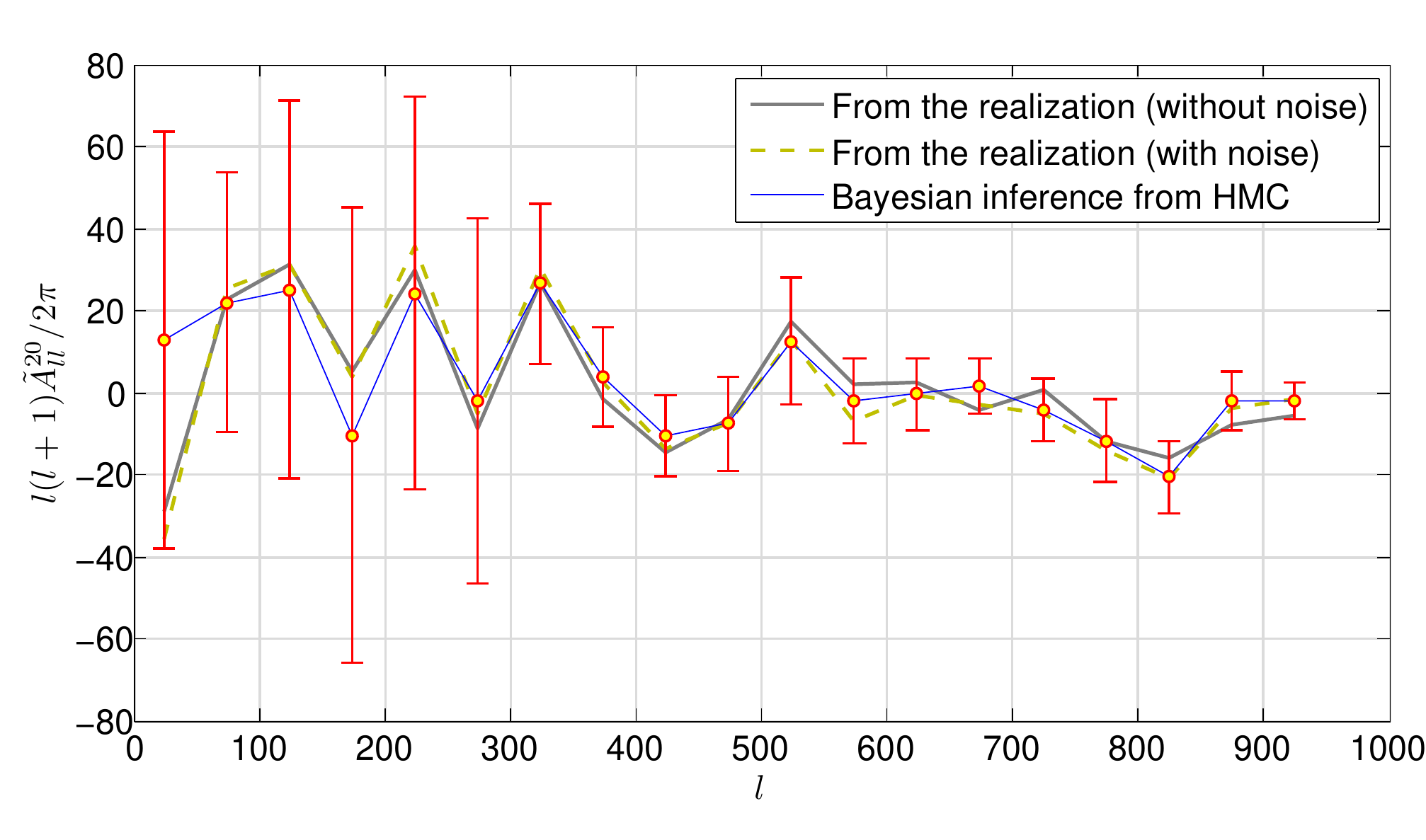}

\includegraphics[width=0.45\textwidth,trim = 0 245 15 255, clip]{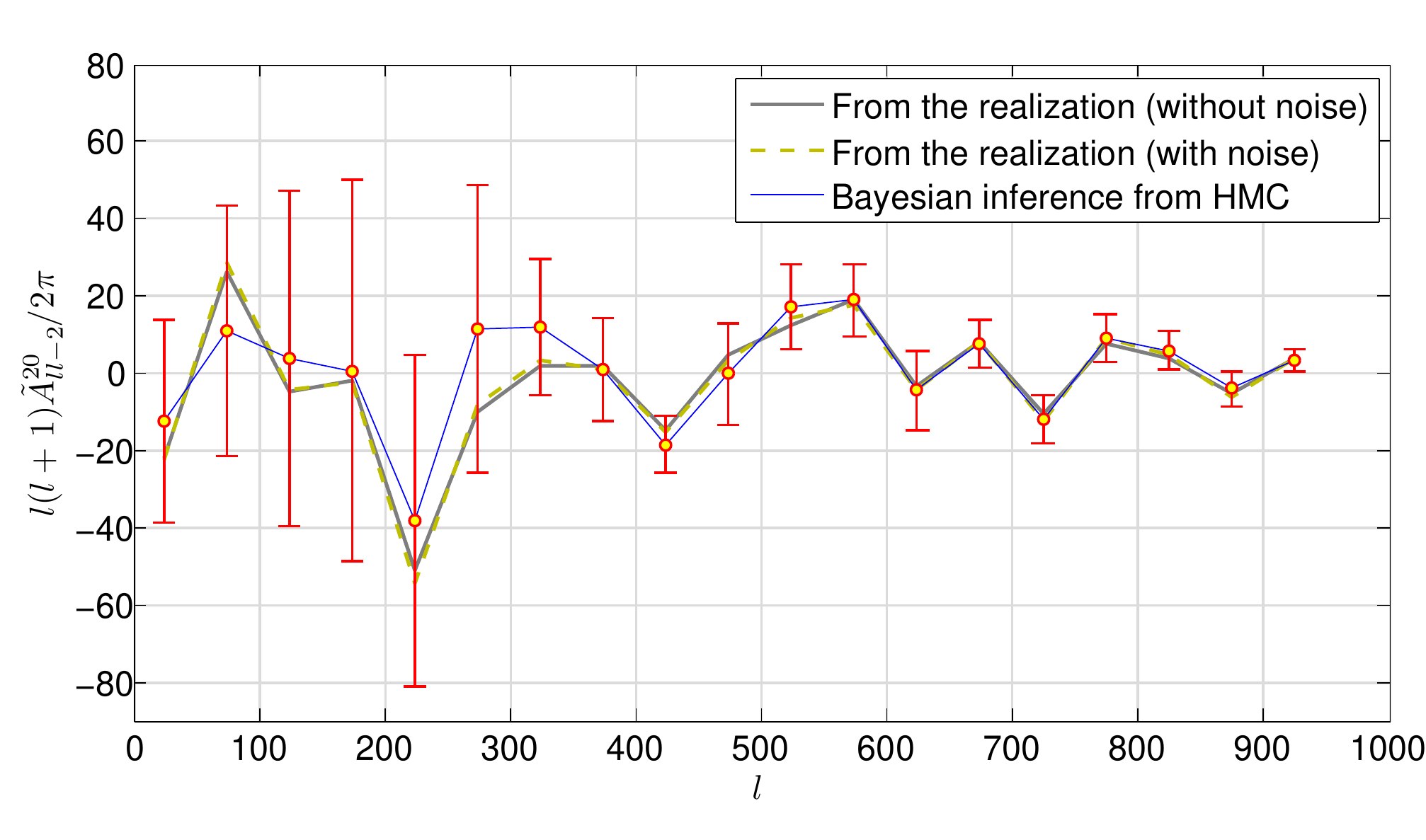}          
\includegraphics[width=0.45\textwidth,trim = 0 245 15 255, clip]{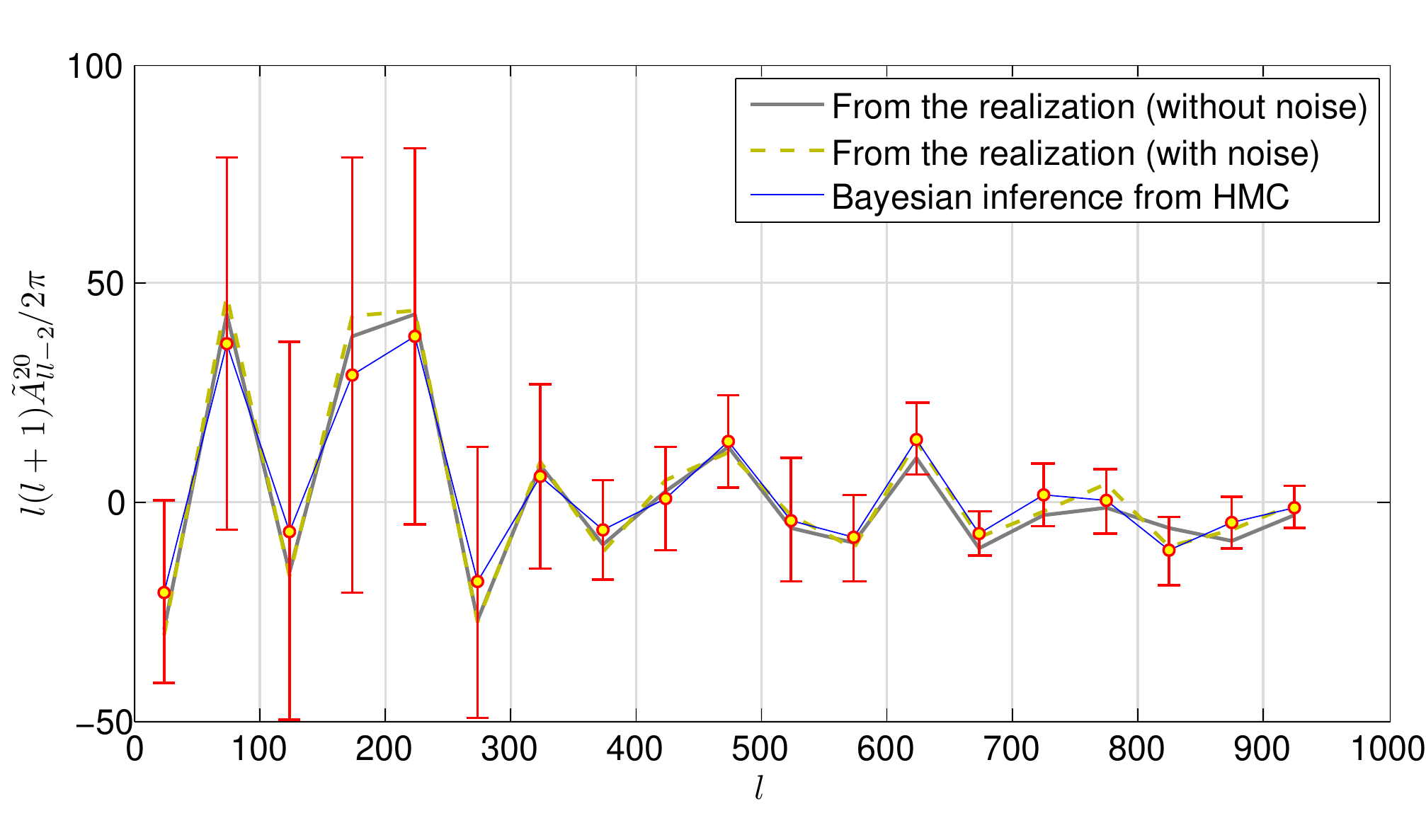}

\caption{\label{fig:Cl_fromNoiseMap} The angular power spectrum and the BipoSH coefficients up to $L=2$ are inferred from 
a statistically isotropic sky map generated using HEALPix. We present the results of our analysis on two different 
realisations in the left and right columns. In the left (right) column, homogeneous, white, Gaussian  random  noise with $\sigma_{n}=10\mu K$
($\sigma_{n}=20\mu K$)  has been added to the signal. In the top row, the solid bold green line shows the input power spectrum for generating the sky map. The thin dark gray line shows the quadratic estimate of the angular power spectrum of the particular realisation before adding noise and the dotted olive line is the same  of the realisation after adding noise. The red data-points show the mean $C_{l}$ recovered using  joint Bayesian inference of the BipoSH 
coefficients up to $L=2$. After the first 10 multipoles we plot the data-points in averages in multipole bins of $\Delta l=20$.
The 2nd, 3rd and 4th row $\tilde{A}^{10}_{ll-1}$, $\tilde{A}^{20}_{ll}$ and $\tilde{A}^{20}_{ll-2}$ are  plots for the BipoSH spectra for $M=0$. The results for $M\neq 0$ are similar and, hence, not plotted here.  We can see that almost all the BipoSH spectra are consistent with zero within $1$ to $2\sigma$, as it should be when the maps are drawn from a statistically isotropic covariance. It should be noted that the vertical scales differ in the different plots.
 }
\end{figure}

\begin{figure}
\centering
\includegraphics[width=0.31\textwidth,trim = 5 260 25 240, clip]{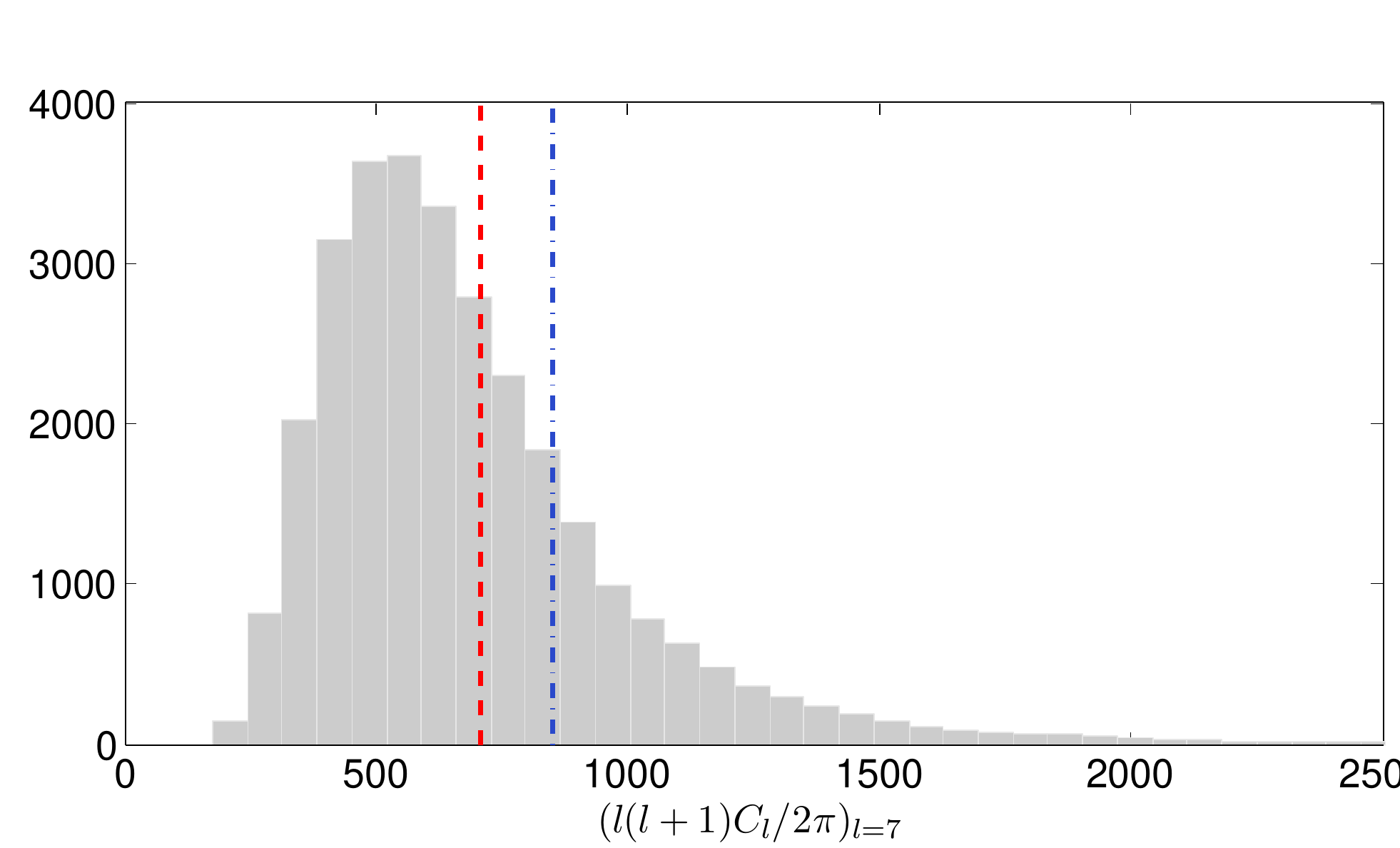}          
\includegraphics[width=0.31\textwidth,trim = 5 260 25 240, clip]{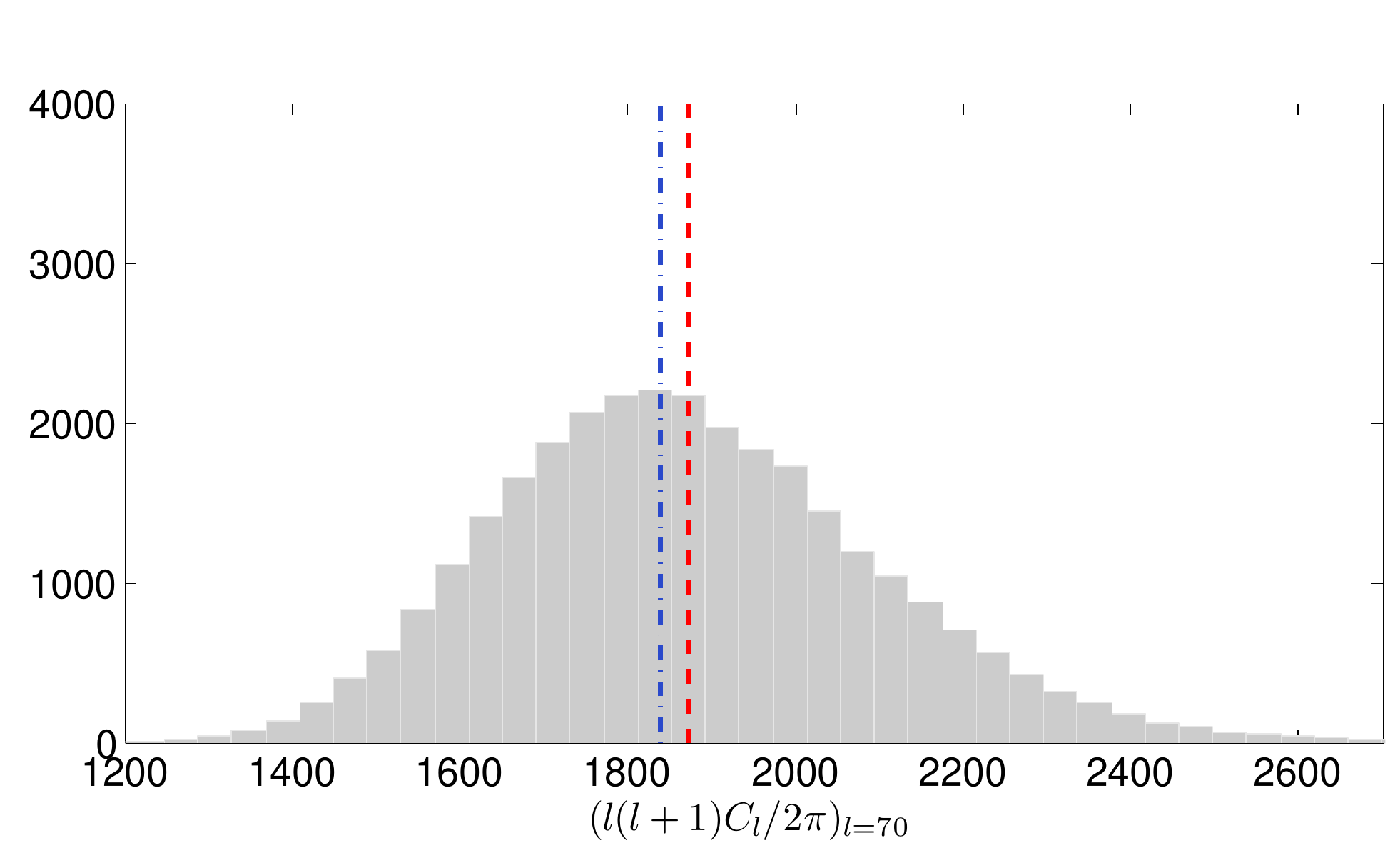}
\includegraphics[width=0.31\textwidth,trim = 5 260 25 240, clip]{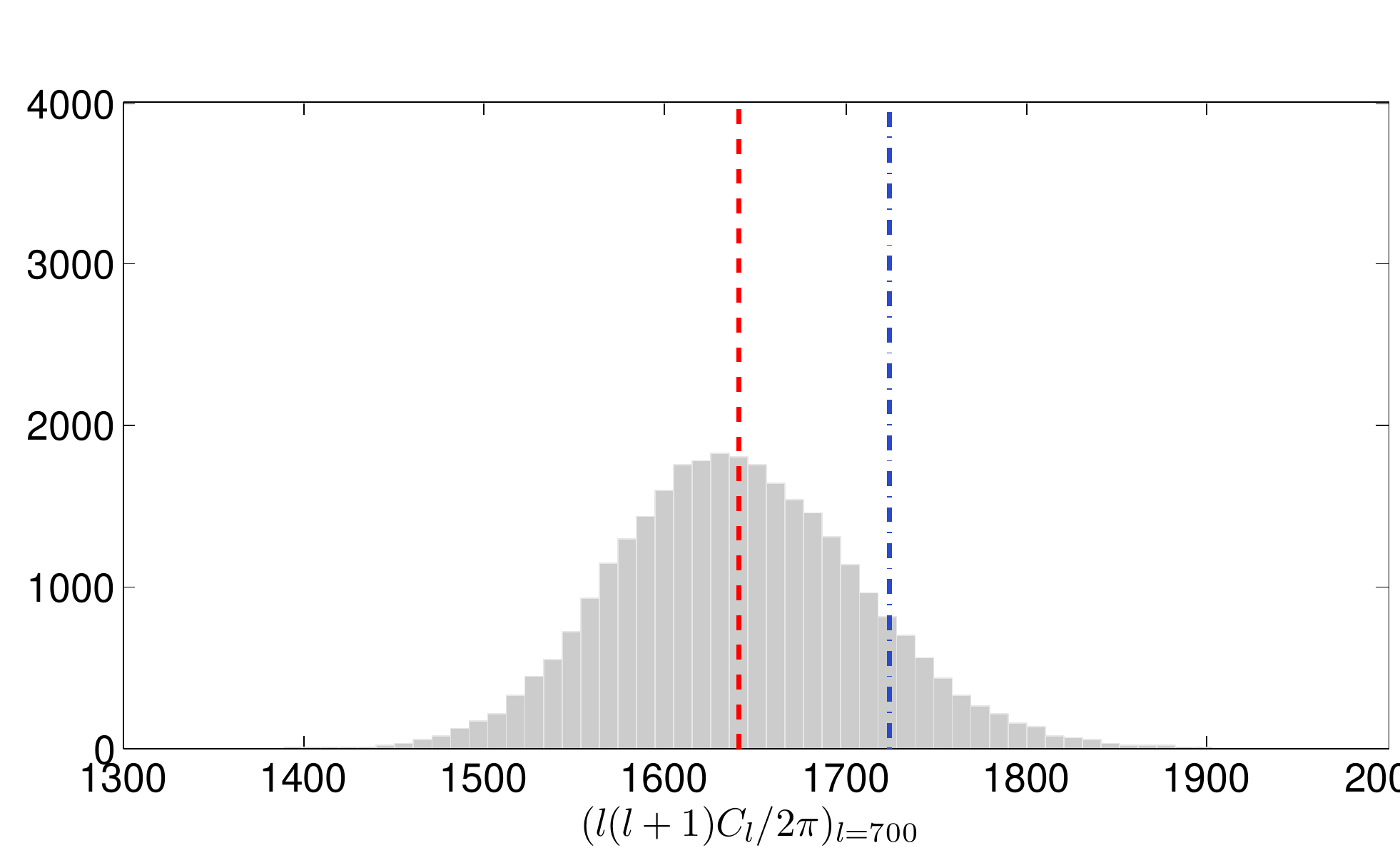}

\includegraphics[width=0.31\textwidth,trim = 5 260 25 240, clip]{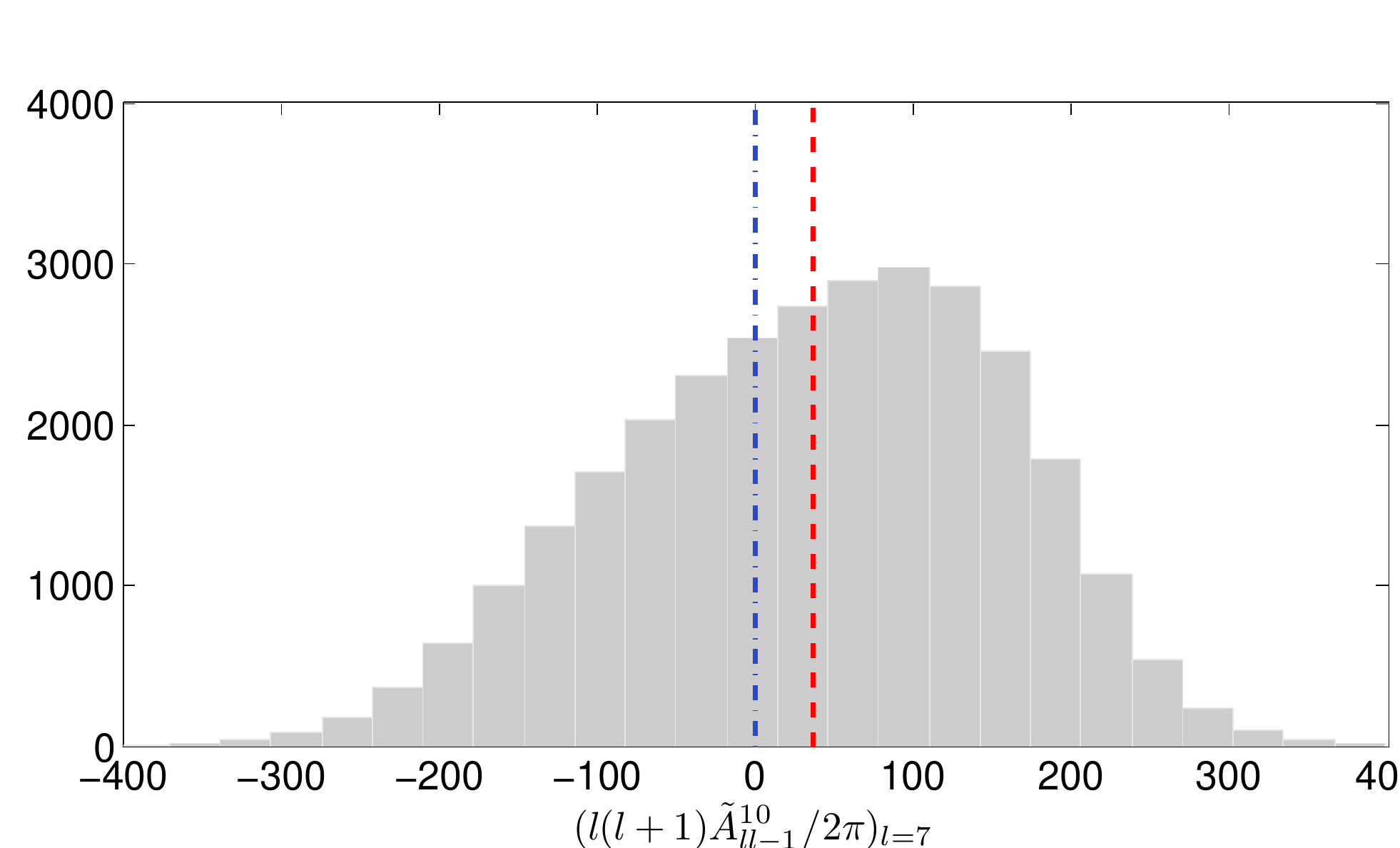}          
\includegraphics[width=0.31\textwidth,trim = 5 260 25 240, clip]{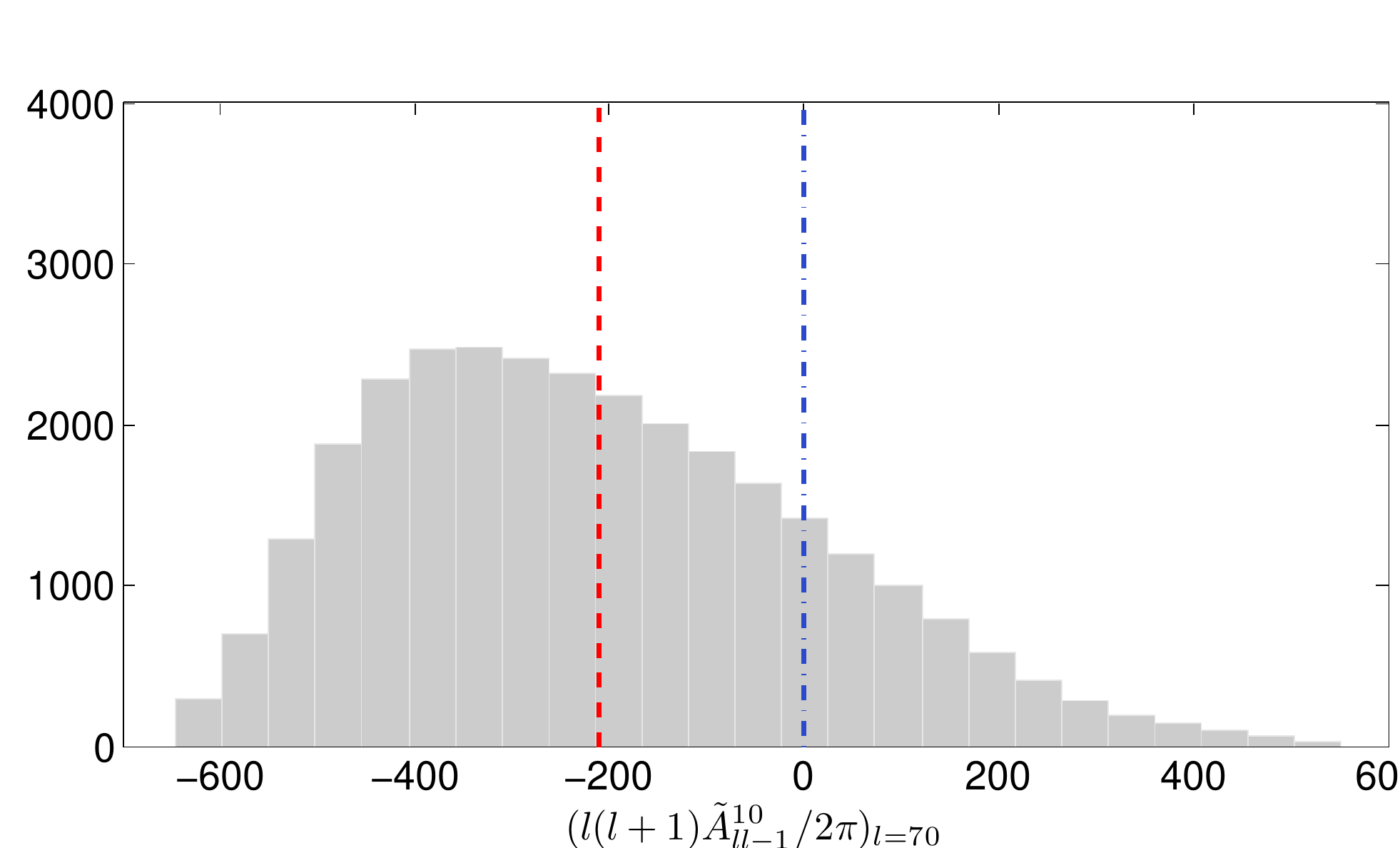}
\includegraphics[width=0.31\textwidth,trim = 5 260 25 240, clip]{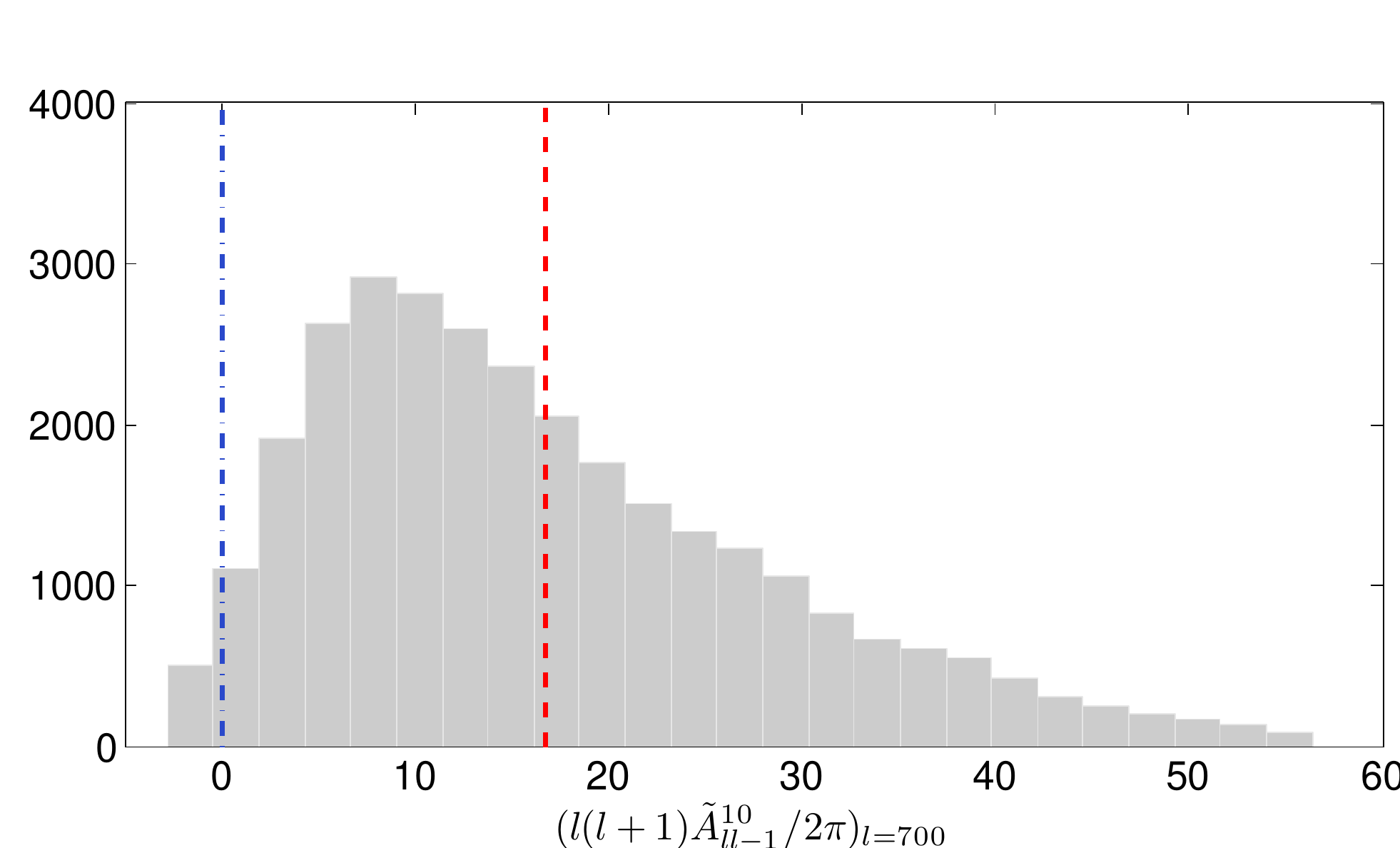}

\includegraphics[width=0.31\textwidth,trim = 5 260 25 240, clip]{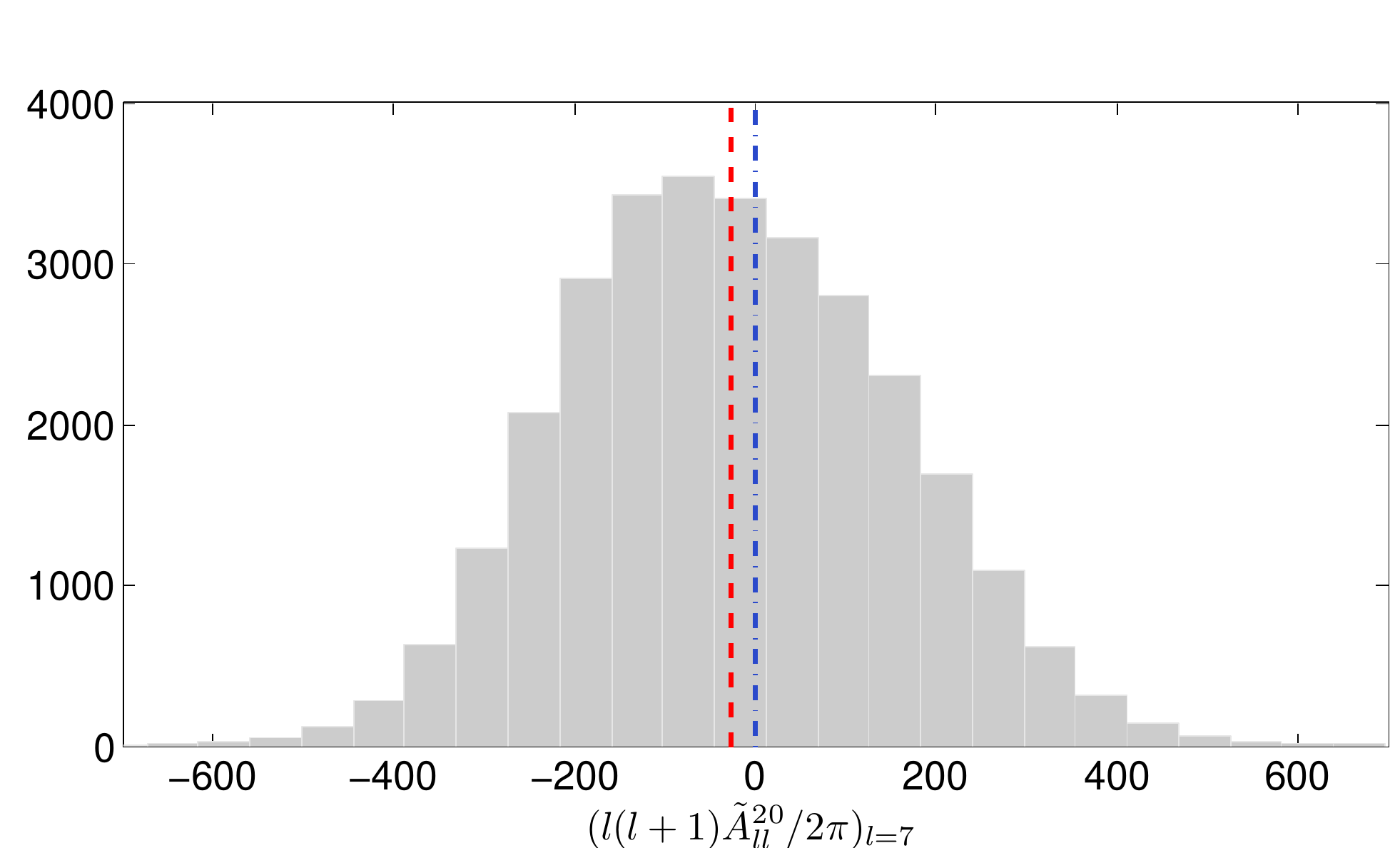}          
\includegraphics[width=0.31\textwidth,trim = 5 260 25 240, clip]{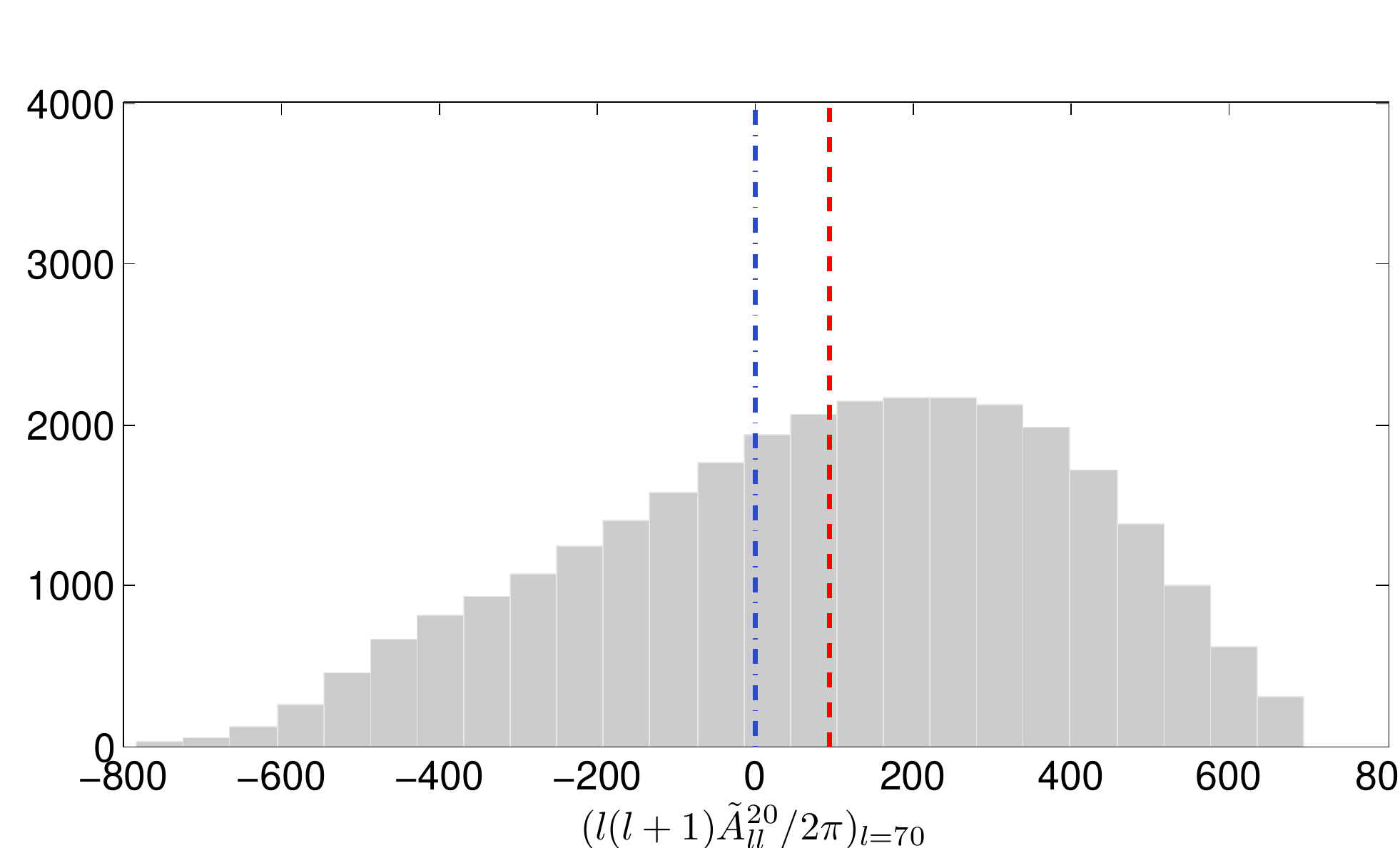}
\includegraphics[width=0.31\textwidth,trim = 5 260 25 240, clip]{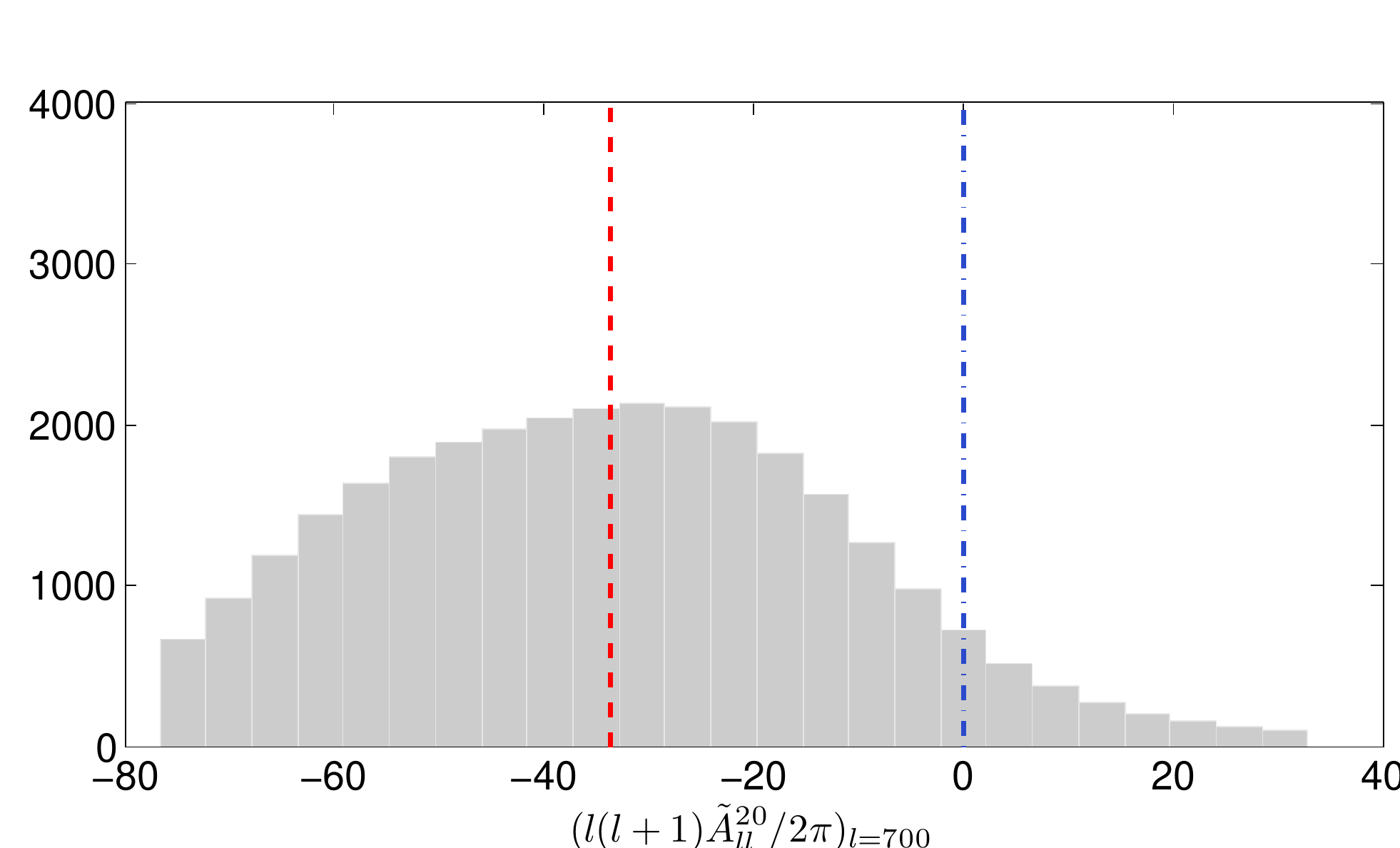}

\caption{\label{fig:ALM_distribution} We show the posterior distribution of $C_{l}$, $\tilde{A}^{10}_{ll-1}$,
$\tilde{A}^{20}_{ll}$  at selected multipoles $l=7,\,70,\,700$ by plotting the number of sample points in different bins. 
Total number of sample points taken is 30,000. 
The posterior distribution is for SI sky map with noise level of $\sigma_{n}=10\mu K$. 
The dashed red  vertical lines  mark  the mean  of  each distribution. Blue dash-dot lines represent the input $C_l$ and the BipoSH 
coefficients.
}
\end{figure}

\section{Computational implementation}
\label{compute}

\subsection{Limitations in $L$ space}
The method, as described above, is applicable to Bayesian  inference of the covariance structure 
that covers arbitrary SI violation. However, the finite resolution and sensitivity of measurements, pose information theoretic 
limitations to the subset of BipoSH spectra that can be expected to be inferred from a single sky map. 
For CMB sky maps that resolve up to a maximum multipole, $l_{\rm max}$, the maximum number of modes of information  available is 
$\sim l_{\rm max}^2$.  Hence, a systematic assessment of all BipoSH spectra ($L_{\rm max}\ge L\ge0$)  seeking to infer 
$\sim L_{\rm max}^3 l_{\rm max}$ coefficients would be information limited beyond  $L_{\rm max} 
\sim l_{\rm max}^\frac{1}{3}$.  
Coarse grained BipoSH spectra recovered in multipole bins  $\Delta l$ will allow a proportionate increase in $L_{\rm max} \sim (l_{\rm max}\Delta l)^\frac{1}{3} $. In case of SI violation studies with  a given model parameterisation of the BipoSH spectra, the number of parameters extracted in this approach together with the angular power spectrum would be similarly limited by some power of $l_{\rm max}$. For example, the SI violation due to weak lensing of CMB is governed by the multipoles $\phi_{LM}$ of the lensing potential field. In this case, one can expect to recover  $\phi_{LM}$, in roughly $\sim l_{\rm max}^\frac{2}{3}$,  independent  bipolar multipole $L$ bins. 

\subsection{Overcoming the time complexity}
There are practical  limitations arising from carrying out the computationally challenging  analysis in reasonable time. 
The first practical issue that we face is related to inverting the covariance matrix. The need to invert the covariance 
matrix is required by  Eqs.(\ref{eq:pdot}) and (\ref{eq:inverse}).  A brute force inversion of the matrix is 
computationally prohibitive.
The inversion of Eq.(\ref{eq:pdot}) can be done by using  Gauss
Seidel method.  Given the expectation that,  $A_{ll'}^{LM},\, L\ne0\wedge M\ne0$,
coefficients are much smaller compared to $A_{ll}^{00}$, the matrix
$S_{lml'm'}$ is a diagonal dominated matrix, making it ideal
for Gauss Seidel method. 

Brute force inversion of $S_{lml'm'}$ in Eq.(\ref{eq:inverse}) is difficult and time consuming except for small lmax (This is an important and interesting sub-case. Many of the anomalies at low multipoles, $l$ can be studied with an $l_{\rm max}$ of a few tens, where brute force inversion is quite possible). However, again using the fact
that, in case of CMB signal the off-diagonal components of the matrix $S_{lml'm'}$ are expected to be much
smaller than the diagonal components dominated by $A_{ll}^{00}$, we can use Taylor series expansion to invert
the matrix. $S_{lml'm'}$ can be decomposed into $S_{lml'm'}=D_{lml'm'}+O_{lml'm'}$,
where $D_{lml'm'}$ is a diagonal matrix consisting only $A_{ll}^{00}$
part of the covariance matrix, i.e.,  $D_{lml'm'}=A_{ll}^{00}C_{lmlm}^{00}\delta_{ll'}\delta_{mm'}$
and $O_{lml'm'}$is the rest of the part of the covariance matrix.
Expanding $\left(S_{lml'm'}\right)^{-1}$  into Taylor
series up to the first order gives us $\left(S_{lml'm'}\right){}^{-1}=\left(D_{lml'm'}+O_{lml'm'}\right)^{-1}=\left(D_{lml'm'}\right)^{-1}-\left(D_{lml'm'}\right)^{-1}O_{lml'm'}\left(D_{lml'm'}\right)^{-1}$.
In realistic case, the $L>0$ BipoSH coefficients $A_{ll'}^{LM}$ being much smaller than $C_{l}$,
this first order approximation works well in the examples studied here.  

Substituting  the expressions for $D_{lml'm'}$ and $O_{lml'm'}$ into  Eq.(\ref{eq:inverse}), we obtain  

\begin{equation}
\partial_{A_{ll}^{00}}\ln\left|S\right|=(2l+1)/A_{ll}^{00}
\end{equation}

\noindent and 

\begin{equation}
\partial_{A_{ll'}^{LM}}\ln\left|S\right|=\left(-1\right)^{L+l+l'+1}\sqrt{(2l+1)(2l'+1)}A_{ll'}^{LM}/\left(A_{ll}^{00}A_{l'l'}^{00}\right)\,.
\end{equation}

\noindent These provide the set equations of motion for $a_{lm}$ and $A_{ll'}^{LM}$ in this approximation. These equations are applicable in case of  weak isotropy violation, which is the case for SI violations in  CMB signal.  In cases where the SI violation signal is strong, i.e., $\left|\frac{A_{ll'}^{LM}}{A_{ll}^{00}}\right|\sim O(1)$, the truncated Taylor expansion approximation used here may not hold. 

\subsection{Stability of numerical integration}

Another computational issue is the choice of the numerical integration method and the mass matrix. 
In  Hamiltonian integrators, though  Leapfrog integrator is common because the integrator preserves 
the Hamiltonian in phase space (symplectic), the propagation error being huge 
we have to use a fourth order symplectic  integrator, namely Forest and Ruth integrator, which performs better and the propagational errors are contained at a manageable level. 

Forest-Ruth algorithm is a combination of three Leapfrog steps and consists of the following steps 
\begin{eqnarray}
x &=& x+\theta \frac{h}{2}v  \nonumber \\ 
v &=& v+\theta h F(x)   \nonumber\\
x &=& x+\theta \frac{h}{2}v   \nonumber \\ 
 \nonumber \\
x &=& x+(1-2\theta) \frac{h}{2}v  \nonumber \\
v &=& v+(1-2\theta) h F(x)    \nonumber \\
x &=& x+(1-2\theta) \frac{h}{2}v  \nonumber \\
 \nonumber \\
x &=& x+\theta \frac{h}{2}v  \nonumber\\
v &=& v+\theta h F(x)   \nonumber \\
x &=& x+\theta \frac{h}{2}v
\end{eqnarray}

\noindent where $h$ is the step size,  $\theta = ({2-\sqrt[3]{2}})^{-1}$ and $x$ represents the variable of integration, i.e., $a_{lm}$ and $A^{LM}_{ll'}$ in our case, and $v$ represents the velocity, i.e.,  $\frac{p_{a_{lm}}}{m_{a_{lm}}}$ and 
$\frac{p_{A^{LM}_{ll'}}}{m_{A^{LM}_{ll'}}}$, respectively. $F(x)$ represents the acceleration, i.e.,  $\frac{\dot{p}_{a_{lm}}}{m_{a_{lm}}}$ and  $\frac{\dot{p}_{A^{LM}_{ll'}}}{m_{A^{LM}_{ll'}}}$. 
It can be seen that the Forest Ruth integrator is a combination of three Leapfrog integrator with step size $\theta h$, $(1-2\theta) h$ and $\theta h$,  respectively. 

Choice of proper mass matrix is crucial for the stability of the integration method. If we can show that each of the Leapfrog step is stable then the entire integration process will also be stable. Here we derive the choice of the mass matrix that ensures that the integration is  stable~\cite{Taylor2008,jain}. 

For the equations  of motion of $a_{lm}$ we have 

\begin{eqnarray}
x_{i+\frac{1}{2}} &=& x_i + \frac{\epsilon}{2} M^{-1}p_{i} \nonumber \\
p_{i+1} &=& p_{i} - \epsilon(S^{-1}+N^{-1})x_{i+\frac{1}{2}}  \nonumber \\
x_{i+1} &=& x_{i+\frac{1}{2}} + \frac{\epsilon}{2} M^{-1}p_{i+1}
\end{eqnarray}

\noindent We ignore $N_{lml'm'}^{-1}d_{l'm'}$ because that part being constant is anyway stable.  
The above equation can be written as 

\begin{equation}
\left[\begin{array}{c}
x_{i+1}\\
p_{i+1}
\end{array}\right]=\left[\begin{array}{cc}
\left(I-\frac{\epsilon^{2}}{2}M^{-1}\left(S^{-1}-N^{-1}\right)\right) & \epsilon M^{-1}\left(I-\frac{\epsilon^{2}}{4}M^{-1}\left(S^{-1}-N^{-1}\right)\right)\\
 -\epsilon \left(S^{-1}-N^{-1}\right)& \left(I-\frac{\epsilon^{2}}{2}M^{-1}\left(S^{-1}-N^{-1}\right)\right)
\end{array}\right]\left[\begin{array}{c}
x_{i}\\
p_{i}
\end{array}\right]
\end{equation}

\noindent For the absolute stability of the integration process we need to ensure that 
the eigenvalues of the matrix are less than unity. The characteristic equation of this matrix 
depends on the choice of the covariance and the mass matrix. Therefore,
if we choose $M=(S^{-1}-N^{-1})^{-1}$ then the characteristic equation will be completely independent of the covariance matrix
and it will be easy to always choose a step size that ensures the  stability of the integration steps. 
However,  in that case $M$ is  a non diagonal matrix that algebraically complicates the computation scheme.
Noting again that $S$ and $N$ can be  expected to be diagonally dominated, we  choose the diagonal approximation,  $M=(\frac{1}{C_{l}}+\frac{1}{N_{l}})^{-1}$, to the  ideal mass matrix. 
In this case the characteristic equation will be nearly  independent
of the choice of the mass matrix making it possible to always choose a step size
for which the integration method stabilises. In our integrator, the step, $h$,  is chosen such that both $\theta h$ and $(1-2\theta)h$ are less than the maximum value of $\epsilon$ that is set as a requirement for the stability of the integration process.

For the analysis and results presented  in this paper, we have assumed the the noise covariance 
matrix to be diagonal in spherical harmonic space, i.e., $N_{lml'm'} = N_l \delta_{ll'}\delta_{mm'}$. However, the choice of the mass matrix will also work for weakly anisotropic noise where the  off-diagonal components of the noise covariance matrix is much smaller then the diagonal components. The case of masked/partial sky  observations can also be addressed by considering a non-SI noise covariance matrix where, in  pixel space, the variance of the noise at  masked pixels  is set to infinity. However,  in this case the choice of the diagonal approximation of the mass matrix may not guarantee the stability of the integration process because the off-diagonal components of the noise covariance matrix in the spherical harmonic space could be comparable to the diagonal components. Therefore, for guaranteed  stability the appropriate choice of the mass matrix  would be non-diagonal and, hence,  algebraically complicate the computational algorithm.  In this paper,  for simplicity,  we restrict to full sky analysis and defer masked sky  analysis  to planned future work  on observed  CMB maps which may necessitate a more complex implementation with non-diagonal mass matrices.

Next we discuss the choice of the mass matrix for the $A_{ll'}^{LM}$, where $L,\, M\ne0$.  In any realistic CMB map 
we can consider that $\left|A_{ll'}^{LM}\right|\ll\left|A_{ll}^{00}\right|$. Therefore, expanding Eq.(\ref{eq:ALMll}) up to first order we 
get similar equation as of Eq.(\ref{eq:pdot}), except that the $(S^{-1}+N^{-1})$ will
be replaced by $\frac{2A_{ll}^{LM}A_{l'l'}^{LM}}{\sqrt{(2l+1)(2l'+1)}}$.
Therefore, following similar logic  as discussed above we choose the mass matrix for $A_{ll'}^{LM}$
as $\left|\frac{\sqrt{(2l+1)(2l'+1)}}{2A_{ll}^{LM}A_{l'l'}^{LM}}\right|$.

The choice of the mass matrix for $A_{ll}^{00}$ is not directly obvious from the above arguments. 
We use the mass matrix of  $A_{ll}^{00}$ as the inverse of its variance, which is consistent with $L=0$ limit of the expression for 
the BipoSH coefficients mass matrix, and is  also found to provide stable integration.


\section{Demonstration of method on simulated CMB sky maps}
\label{results}

In this section we demonstrate  our Bayesian inference method on some representative examples. We consider  a variety of  simulated CMB maps, such as, statistically isotropic sky map, 
non-SI map with a non-circular beam  signal as detected in WMAP-7 year data~\cite{wmap7-anomalies}, non-SI map with dipole modulation signal and the Doppler boosted signal on SI violation 
as detected with  Planck observations~\cite{planck_isotropy,Planck2013XXVII,Ade2015XVI}. For all these cases,  the method is tested at different noise levels. In presenting  the results,  we plot the BipoSH spectra 
$\tilde{A}_{ll'}^{LM}=\frac{\sqrt{2L+1}}{\sqrt{2l+1}\sqrt{2l'+1}}\frac{1}{\mathcal{C}_{l0l'0}^{L0}}A_{ll'}^{LM}$, as  defined for even parity
BipoSH coefficients with even value of $L+l+l'$.

\subsection{Statistically Isotropic CMB map}

First we test the method on  statistically isotropic sky maps. We produce 
SI map using HEALPix~\cite{healpix} at the resolution $N_{{\rm side}}=512$ pixelisation. Being
statistically isotropic, all the BipoSH spectra, except $\tilde A_{ll}^{00}$, are expected to statistically consistent with zero. We then add SI 
Gaussian random, white, noise with zero mean and standard deviation $\sigma_n$ to the signal map realisation.  We apply  our algorithm 
for joint Bayesian inference of  $C_l$ and the other BipoSH coefficients up to $L\le2$. 

The results are shown in Fig.~\ref{fig:Cl_fromNoiseMap}. We test
the algorithm at two different noise levels, $\sigma_n=10\mu K$ and  $\sigma_n=20\mu K$.
For plots of $C_l$  we  adopt a hybrid axis which logarithmic up to multipole, $l=100$ and linear beyond. Also, we plot the individual 
mean values at each multipole, $l\le 10$, and provide band power average in multipole bin size of $\Delta l=20$ at larger multipoles. 
The plots show that the algorithm  recovers the input  power spectrum perfectly  up to the high multipoles at both the noise levels. We show the BipoSH coefficients,  $\tilde{A}^{10}_{ll-1}$, $\tilde{A}^{20}_{ll}$ and 
$\tilde{A}^{20}_{ll-2}$ in the 2nd, 3rd and 4th column of Fig.~\ref{fig:Cl_fromNoiseMap}. 
The values of these BipoSH coefficients are consistent with zero within $1$ to $2\sigma$. 
Though we do not present plots for  the other $M\neq 0$ BipoSH coefficients (such as $\tilde{A}^{11}_{ll-1}$,$\tilde{A}^{21}_{ll}$ etc. ), 
we   verify that they are also consistent with zero within $1$ to $2\sigma$. 

In Fig.~\ref{fig:ALM_distribution}  we show the marginalised probability distribution of $C_{l}$, $\tilde{A}^{10}_{ll-1}$,
$\tilde{A}^{20}_{ll}$  for the multipoles $l=7,\,70,\,700$ respectively for the analysis with  $\sigma_n=10\mu K$ noise level. 
In the absence of noise, the inverse of the angular power  $C_l$ is known to follow the  $\Gamma$-distribution
that  tends to Gaussian at high multipoles. Our analysis shows similar behaviour for the recovered posterior distributions. The  
theoretical distributions of the  $\tilde{A}^{10}_{ll-1}$ and $\tilde{A}^{20}_{ll}$ are not known, although, some exploratory study has been carried out
in literature~\cite{Joshi2012a}.

\begin{figure}
\centering
\includegraphics[width=0.49\columnwidth,trim = 10 260 10 260, clip]{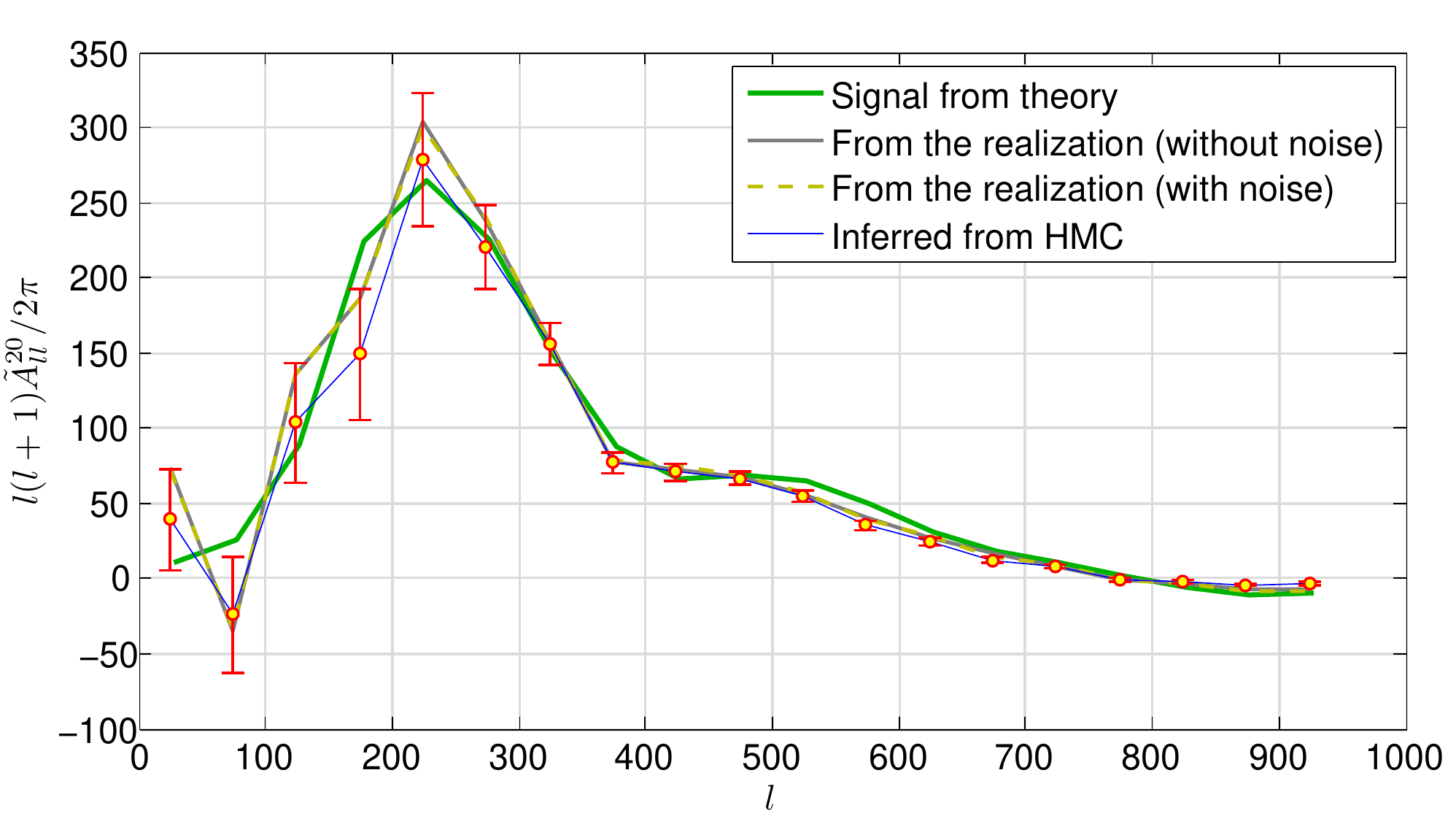}
\includegraphics[width=0.49\columnwidth,trim = 0 260 10 260, clip]{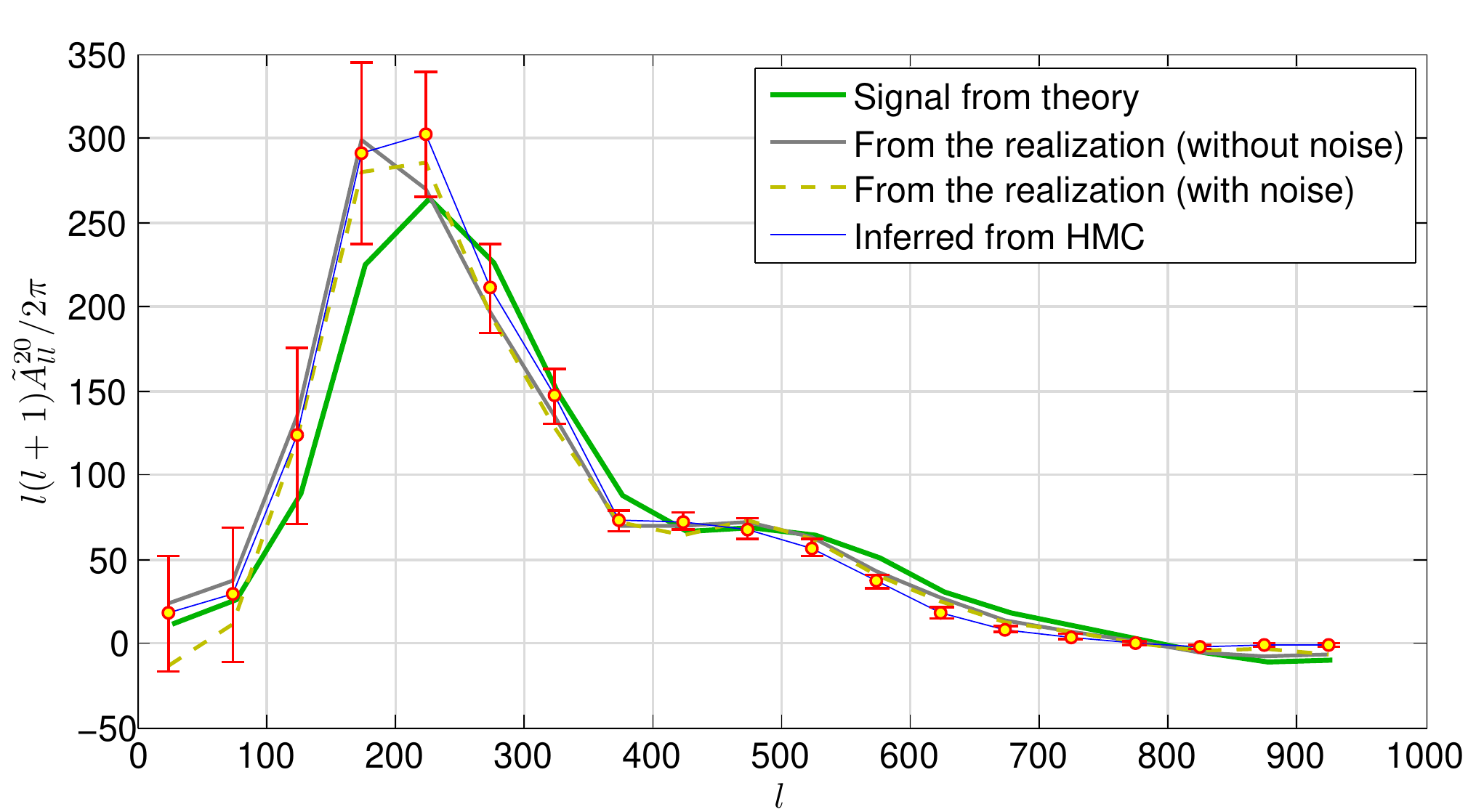}

\includegraphics[width=0.49\columnwidth,trim = 0 260 10 250, clip]{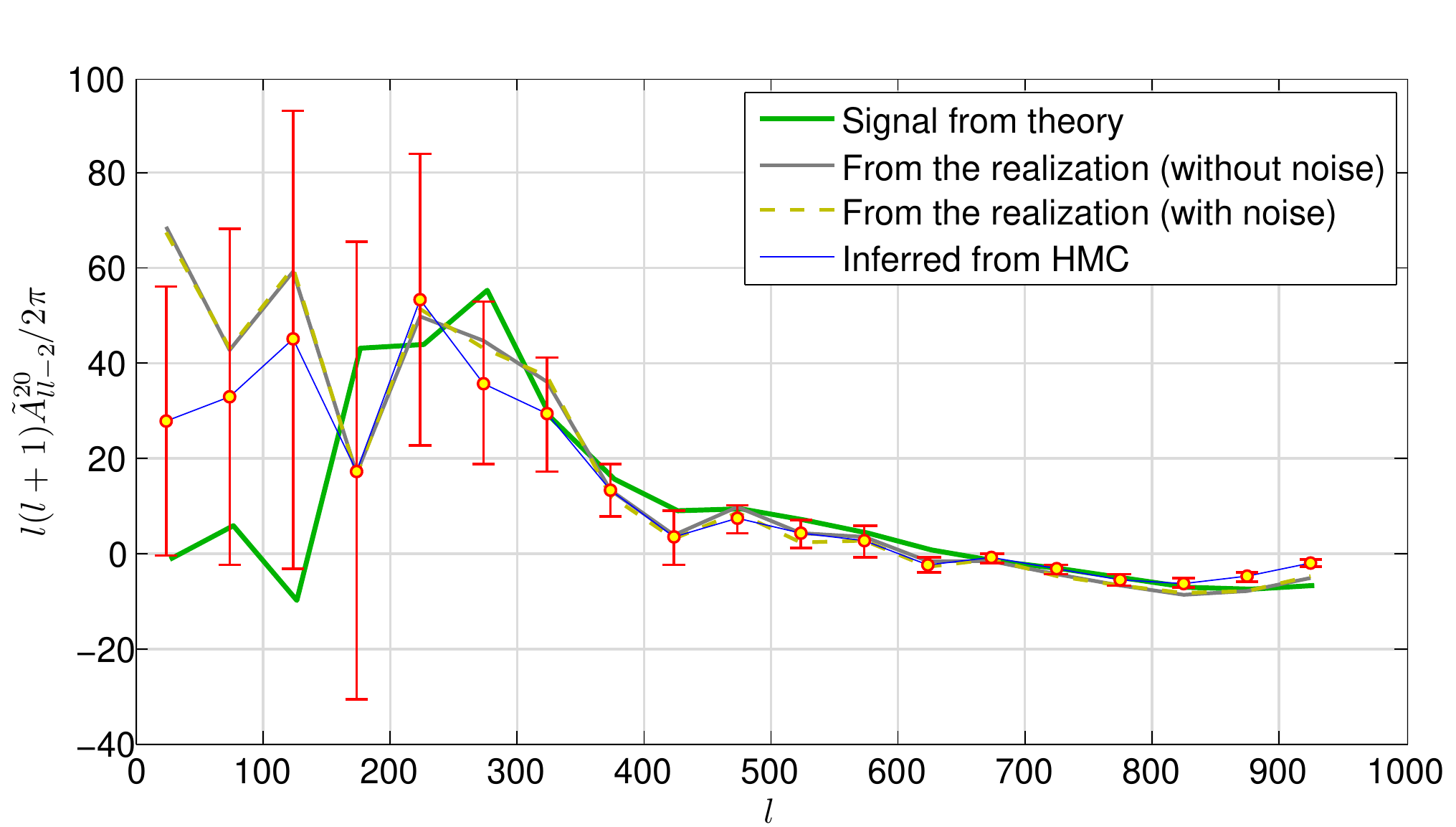}
\includegraphics[width=0.49\columnwidth,trim = 0 260 10 250, clip]{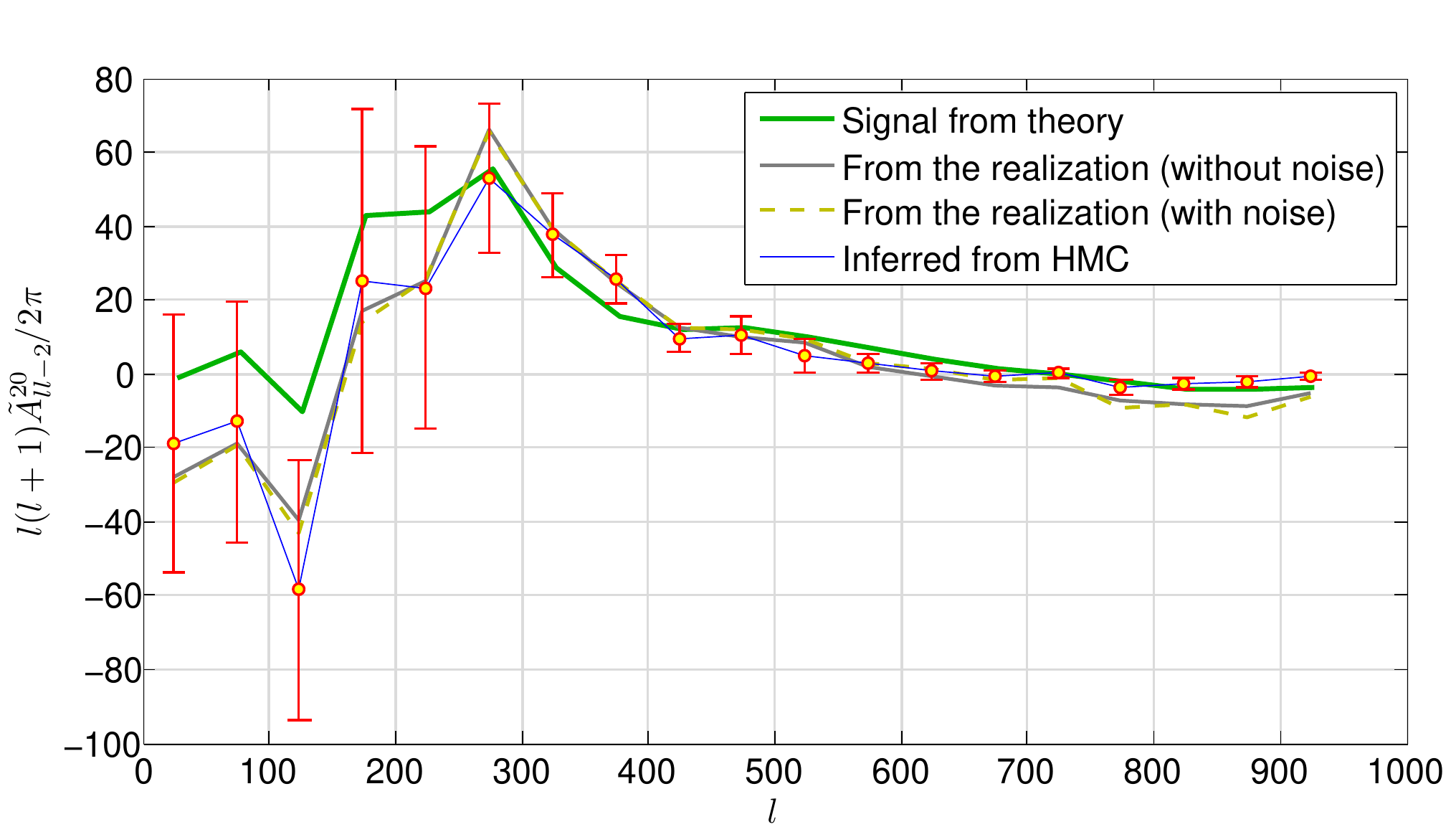}

\caption{\label{fig:A20ll2scannedmap}
BipoSH spectra, $\tilde{A}_{ll}^{20}$ and $\tilde{A}_{ll-2}^{20}$, inferred  from
a non-SI  sky map generated by scanning the sky using WMAP-9 beam
and scan strategy. The dark grey plot shows the quantities obtained from  a direct quadratic estimation  
from the original sky map and olive dashed curve shows the same quantities  after adding noise with the realization. 
The green plot is the theoretical BipoSH values estimated by averaging the BipoSH samples from 30 simulations. 
Blue curve with red error-bars show the recovered value of the same from Bayesian inference. 
The left column corresponds to a noise level of 
$\sigma_{n}=10\mu K$ and the right column to a noise level of $\sigma_{n}=20\mu K$.ONote that the vertical scales are different in the different plots.
}
\end{figure}

\subsection{Beam anisotropy sourced non-SI map}

WMAP-7  CMB map showed significant detection of SI violation signal  in $\tilde{A}^{20}_{ll}$ and 
$\tilde{A}^{20}_{ll+2}$~\cite{wmap7-anomalies}. Later it was revealed that the signal SI violation signals detected 
 were caused due to the mild non-circularity of  the beam response of the WMAP experiment coupled to its 
particular scan pattern~\cite{Das2014}. Therefore  as a non-SI case study,  we take a SI  HEALPix 
map and scan it  with a WMAP-like  scan strategy using  the publicly available (mildly non-circular symmetric) WMAP W1 band beam response maps  to generate  a simulated time ordered  data stream.  From the time ordered data  we reconstruct the map~\cite{Das2013}.  This  procedure introduces SI violation signatures ($\tilde{A}_{ll}^{20}$ and $\tilde{A}_{ll+2}^{20}$) in the scanned map~\cite{Das2014,Joshi2012}. We then add white Gaussian noise of amplitude either $10\mu K$ and $20\mu K$ to the maps. These noisy maps are then used to infer the BipoSH spectra.

The BipoSH spectra of $\tilde{A}_{ll}^{20}$ and $\tilde{A}_{ll+2}^{20}$ recovered
from the analysis are plotted in  Fig.~\ref{fig:A20ll2scannedmap}. The plot shows
that the BipoSH spectra is very well recovered up to a very high $l$. 
Recovery of $\tilde{A}^{20}_{ll}$ is very good at both the noise levels. We note that  $\tilde{A}^{20}_{ll-2}$, however, 
mildly  deviates from the input  signal at high multipoles, $l>800$.

\begin{figure}[h]
\centering
\includegraphics[width=0.60\columnwidth,trim = 0 220 10 220, clip]{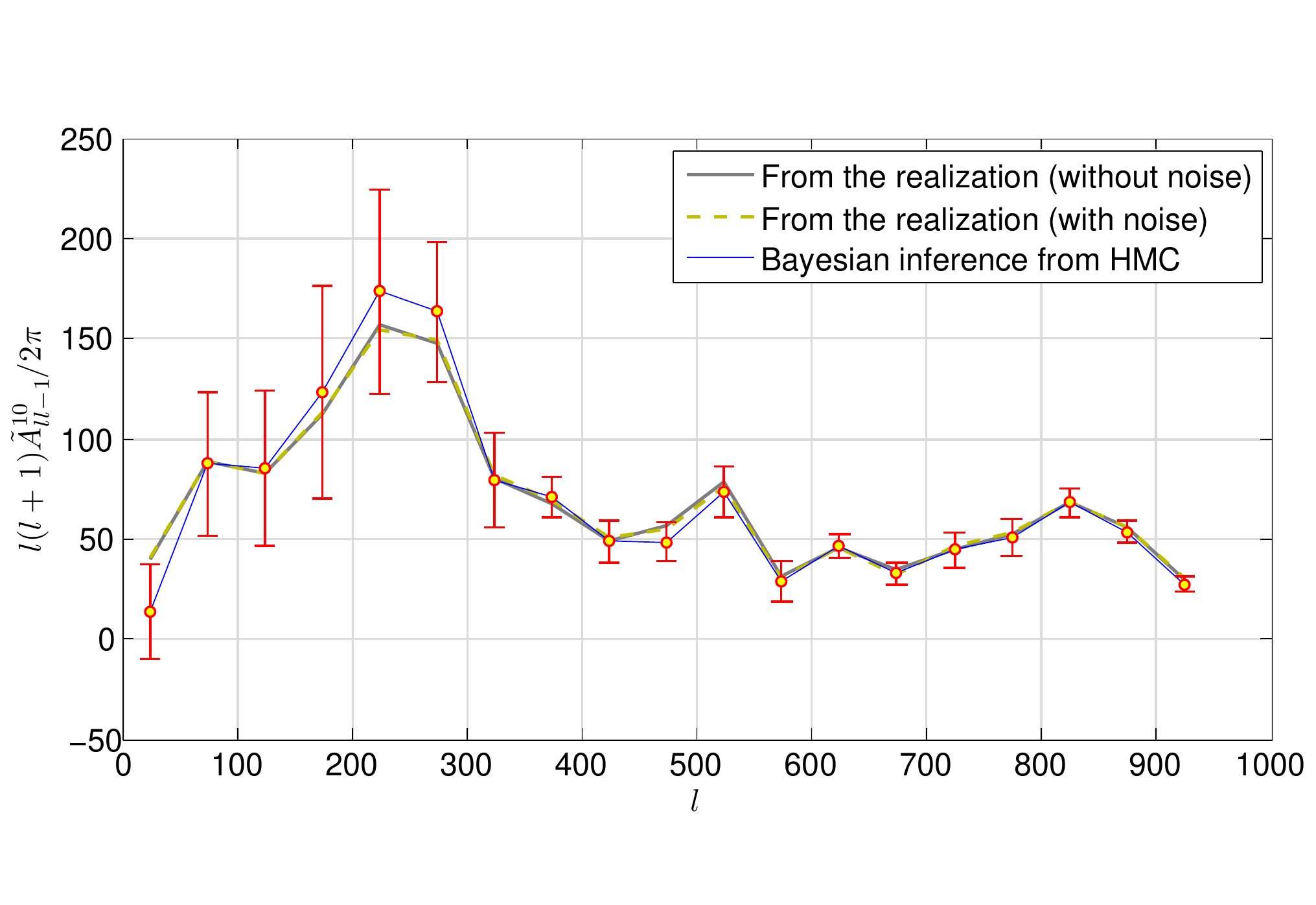}

\caption{\label{fig:Dipolemodulation}$\tilde{A}_{ll+1}^{10}$ estimated from
a non-SI  skymap generated multiplying a SI map with $(.95+0.5 T_d)$, where $T_d$ is a dipole modulation 
oriented along the $\hat z$ direction.  The light gray and green plot show the $\tilde{A}_{ll+1}^{10}$ quadratic estimations
from the original skymap  before and after adding noise. Blue curve with red error-bars shows the
recovered value of the same from our inference. For this analysis we use $\sigma_{n}=10\mu K$.}
\end{figure}

\subsection{Dipolar modulation non SI map}

In  the second non-SI case study,  we consider  a dipole modulated sky map. We generate a SI skymap $T(\gamma)$ using HEALPix 
and then multiply it with $(.95+0.5 T_d)$ where $T_d$ is a dipole modulation oriented along $z$ direction. The map, thus generated,  
has explicit  SI violation  signals captured in $\tilde{A}^{10}_{ll-1}$. Then we add  instrumental noise at $\sigma_{n}=10\mu K$
to the map. From this map we recover the signal in $\tilde{A}^{10}_{ll-1}$. The results are shown in Fig.~\ref{fig:Dipolemodulation}.
It can be seen that the recovery is good at all the multipoles. 

\subsection{Non-SI inference in the context of a physical model: Doppler boost}

\begin{figure}[h]
\centering
\includegraphics[width=0.46\columnwidth,trim = 30 230 0 230, clip]{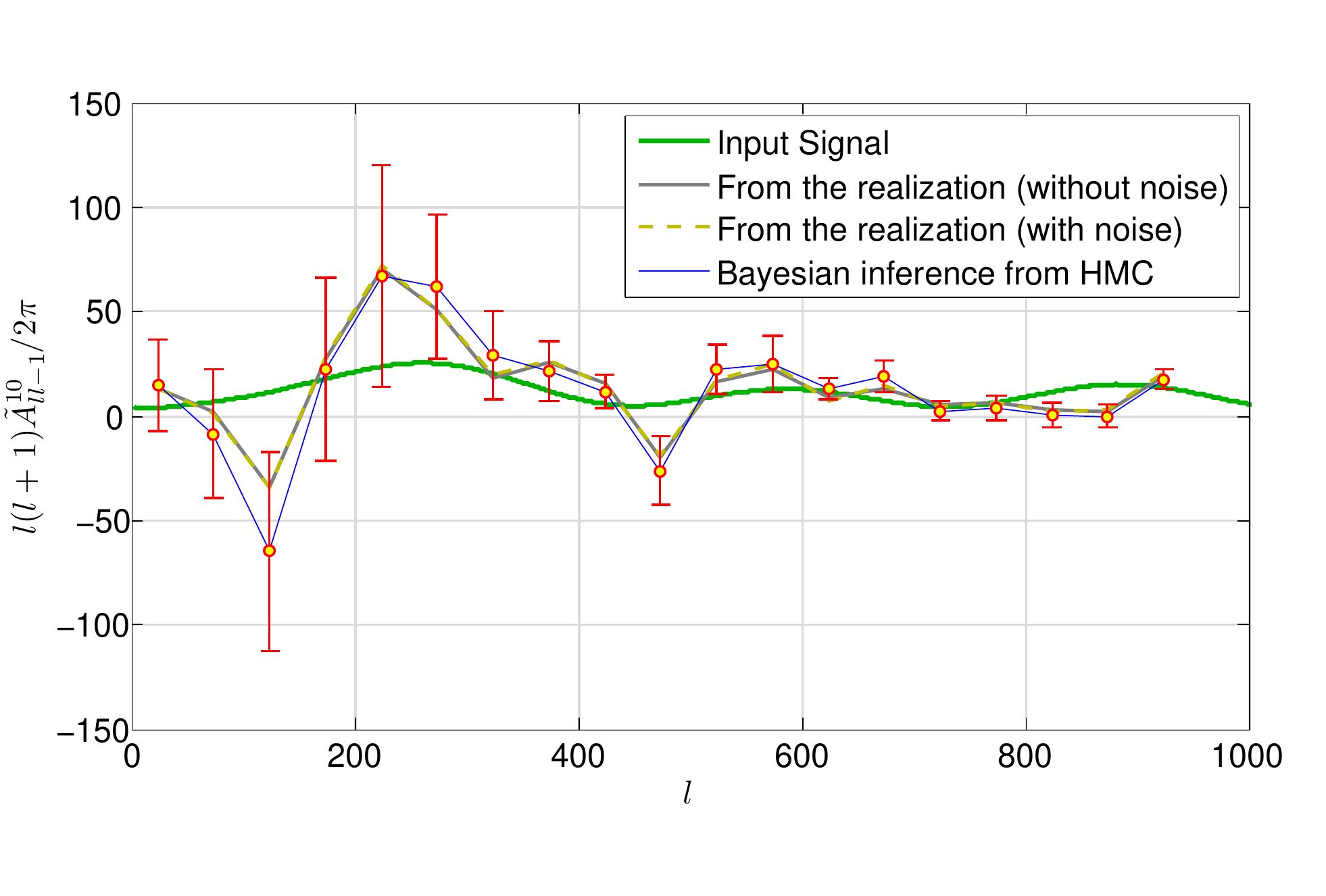}
\includegraphics[width=0.49\columnwidth,trim = 0 230 0 230, clip]{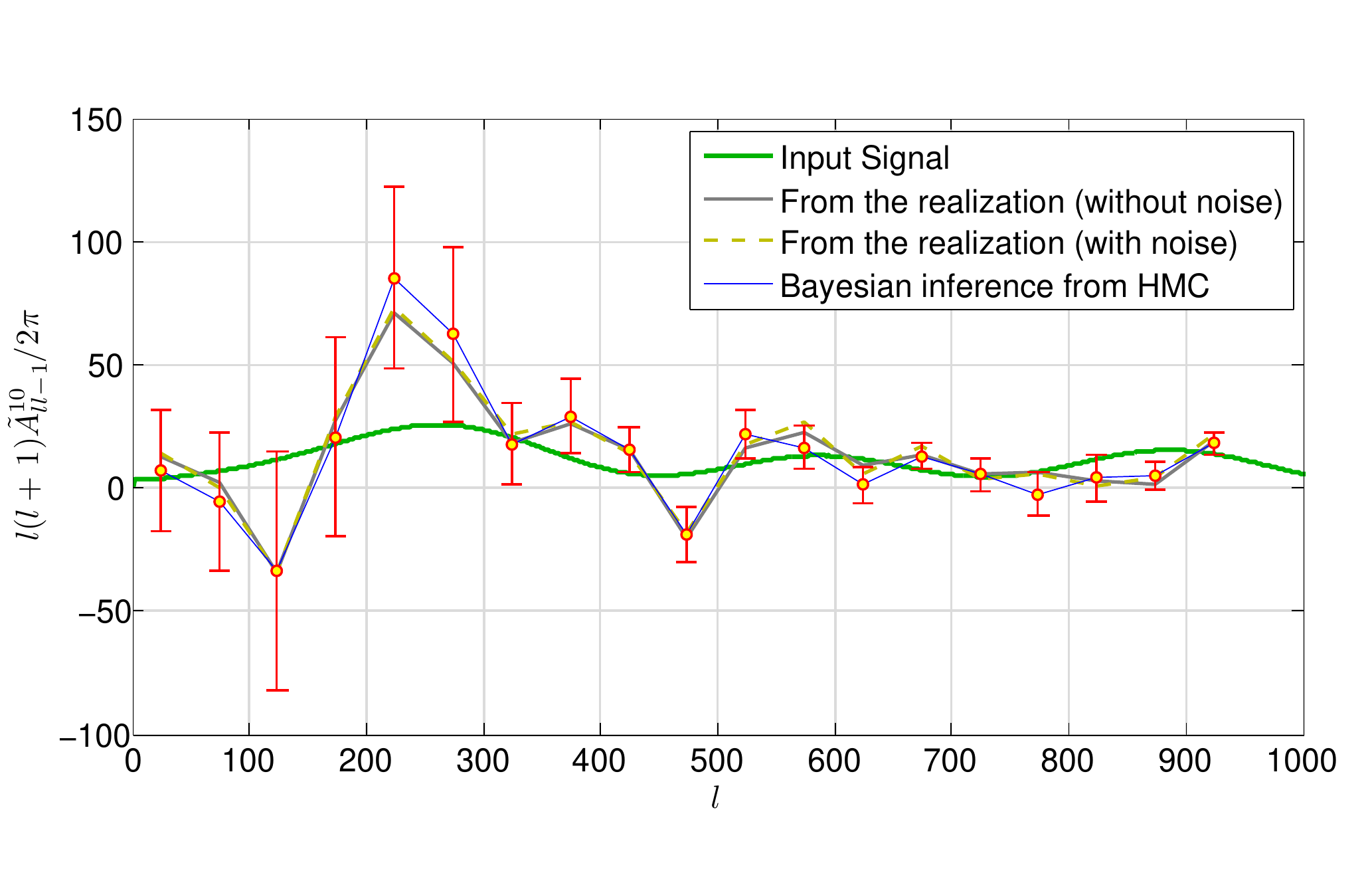}


\caption{\label{fig:Normalized_A10ll1} The BipoSH $\tilde{A}_{ll+1}^{10}$ estimated from
a SI violation arising from Doppler boost along $\hat z$ direction. The non-SI signal map is generated using CoNIGS. 
The dark green plot is the input signal. Light green and grey plot show the $\tilde{A}_{ll+1}^{10}$ from the 
original sky map  before and after adding noise. 
Blue curve with red error-bars shows the recovered value of our HMC based inference. Left plot is for $\sigma_{n}=10\mu K$
and the right plot is for $\sigma_{n}=20\mu K$. The recovery of $\tilde{A}_{ll+1}^{10}$ is accurate up to very high $l$. 
}
\end{figure}

Due to the motion of our galaxy with respect to the CMB rest frame the observed CMB signal experiences Doppler boost. Apart from boosting the monopole temperature, the Doppler boost also  affects the CMB temperature fluctuations. Its effect on the temperature fluctuations are two fold. Firstly, it produces a modulation effect, that amplifies the temperature along the velocity direction and reduces in the opposite direction. Second effect is the relativistic aberration effect that squashes the anisotropy pattern on one side of the sky and
stretches it on the other, effectively mixing the angular scale. As a result a specific form of SI violation is introduced in the CMB maps. Planck 2013 results measured the non-SI  signal 
associated with effect of the Doppler boost~\cite{planck_isotropy,Planck2013XXVII}.  
 In the final non-SI case study, we generate a non-SI sky map consistent with the 
 signal from the Doppler boost along the $\hat z$ direction using CoNIGS~\cite{CoNIGS}.
As earlier, we  add  Gaussian white noise with amplitude $10\mbox{\ensuremath{\mu}K}$ and $20\mu K$ and run our analysis
on the noisy maps thus produced to recover the BipoSH spectra. 
The results of the recovery of the relevant  $A^{10}_{ll+1}$ BipoSH spectra  is shown in Fig.~\ref{fig:Normalized_A10ll1}. 
The figure  shows the values of  $\tilde{A}_{ll+1}^{10}$ recovered from a particular realization at 
two different noise levels. For both noise levels, the recovered values match very well with those estimated from  the input map. 
The recovered means are plotted as  plotted as band power  averages in multipole bins, $\Delta l= 50$.

In this particular known cause of SI violation, the non-SI signal in BipoSH is very simply  related to  the Doppler boost vector, $\vec\beta={\vec v}/{c}$ related to our peculiar motion with respect to the CMB rest frame.. Hence, as an illustration of  non-SI parameter inference readily possible in our methodology, we  carry out  a Bayesian inference of the  boost parameter, $\beta^{1M}$, where 
$\beta^{1M}=\int\beta Y^{*1M}d\Omega$ is the spherical harmonic 
decomposition of $\beta=\frac{v}{c}$. 



\begin{figure}[h]
\centering
\includegraphics[width=0.32\columnwidth,trim = 10 230 20 250, clip]{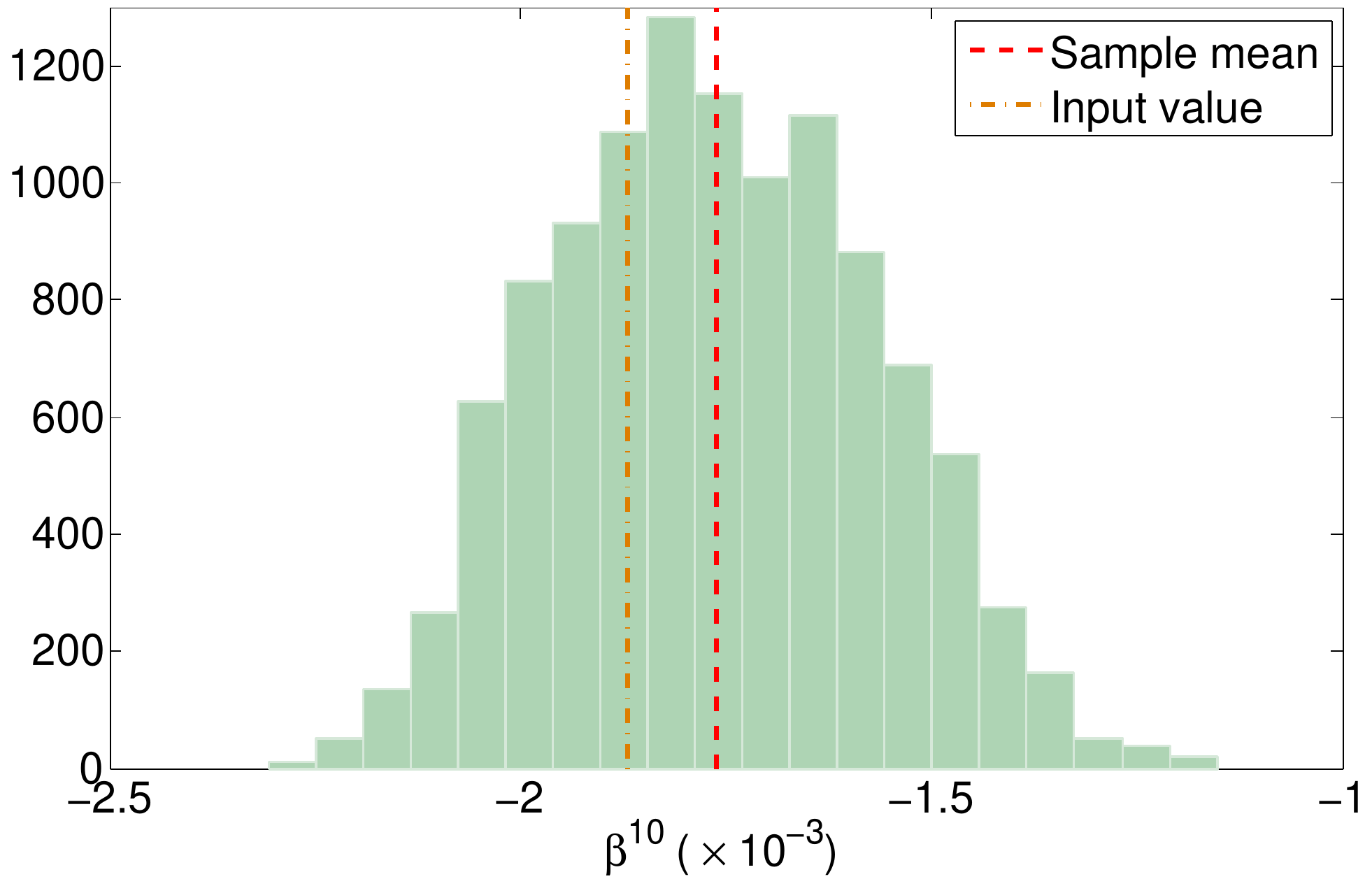}
\includegraphics[width=0.32\columnwidth,trim = 10 230 20 250, clip]{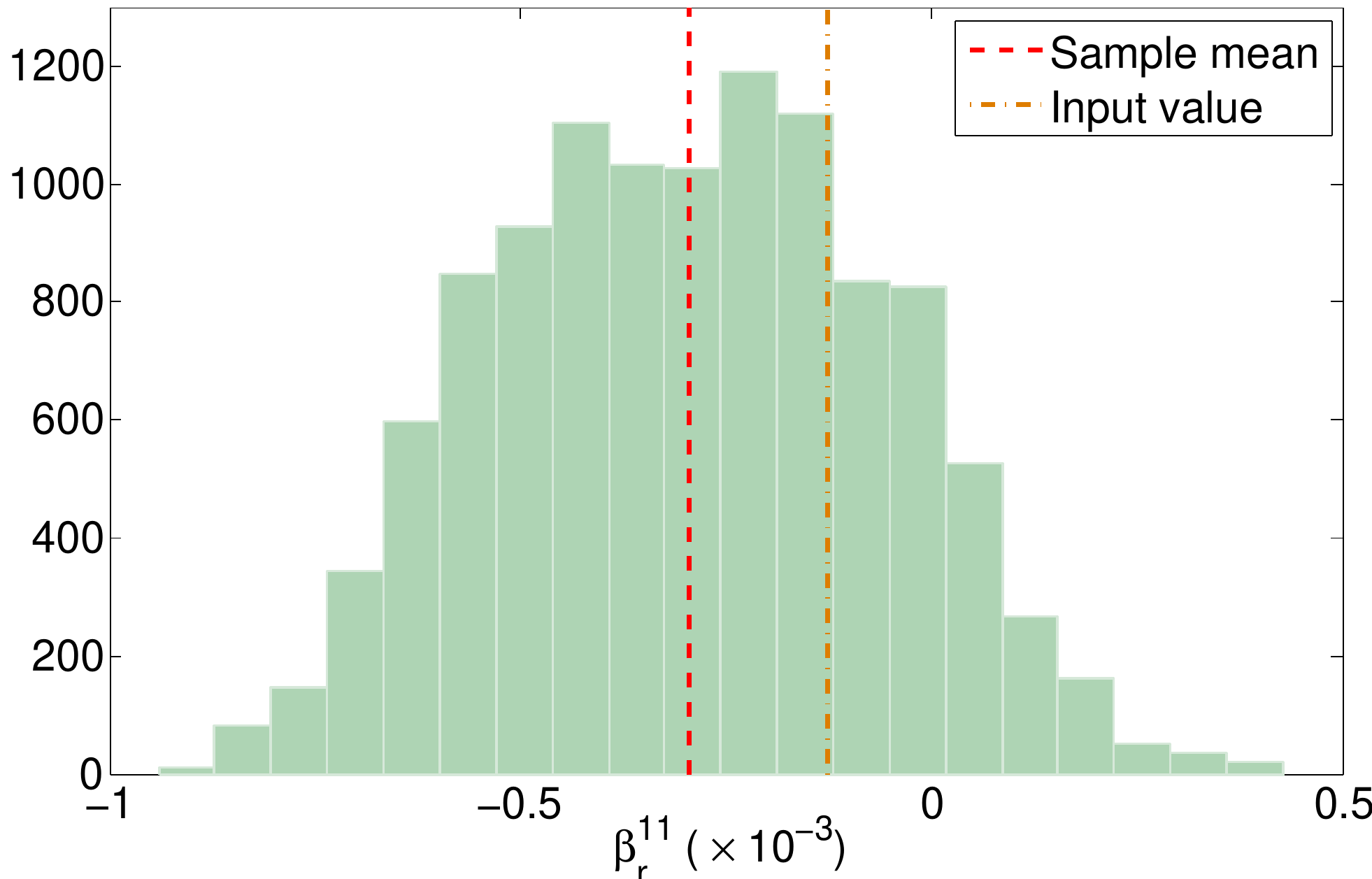}
\includegraphics[width=0.32\columnwidth,trim = 10 230 20 250, clip]{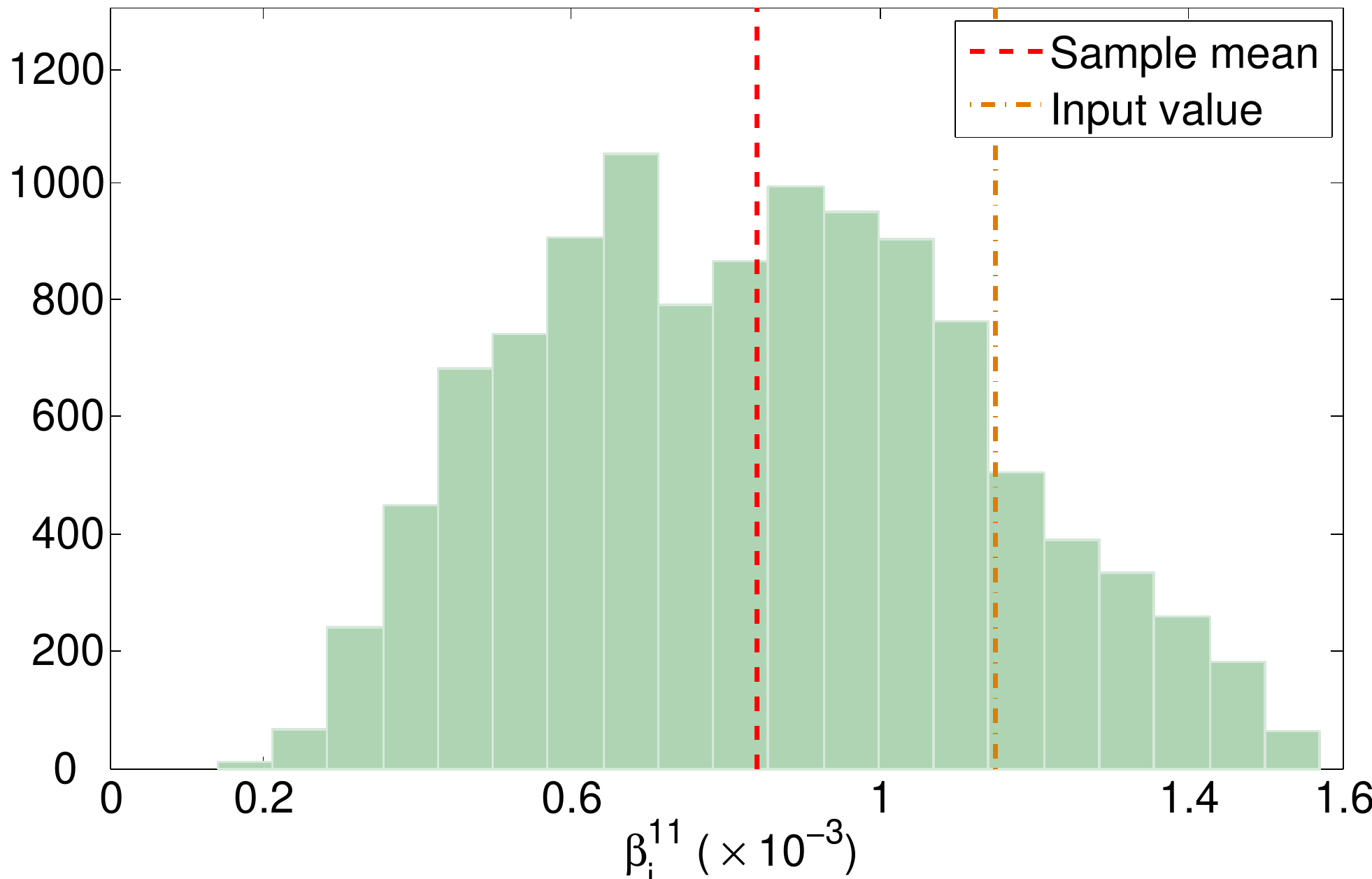}

\caption{\label{fig:beta1} We calculate the posterior of $\beta^{1M}$ from a realization generated from input 
$\beta^{10} = - 1.87\times 10^{-3}$  and $\beta^{11} = - 1.24\times 10^{-4} + 1.18\times 10^{-3}i$, by directly 
sampling the likelihood. From the recovered posterior, we obtain the mean 
$\beta^{10} = - 1.76\times 10^{-3}$  and $\beta^{11} = - 2.94\times 10^{-4} + 0.84\times 10^{-3}i$ The sample map has
an isotropic Gaussian random noise $\sigma_n=20\mu K$. We generate 11,000 samples which are distributed in 20 bins in 
each of the plot. The injected value of $\beta^{1M}$ and the sample mean from our analysis are also marked by vertical lines in the graph.
}
\end{figure}

The posterior of $\beta^{1M}$ can be obtained by directly sampling the $\beta^{1M}$ from the probability 
distribution $P(S_{lml'm'},a_{lm}|d_{lm})$ and considering that the only SI violation signal in the map originates from the Doppler boost. Under such assumption $A^{1M}_{ll'}=\beta^{1M}H^M_{ll'}$ 
and all other BipoSH coefficients are $0$. The equation of motion for $\beta^{1M}$ can be obtained as 

\begin{eqnarray}
\dot{p}_{\beta^{1M}} = \frac{\partial \ln P(S_{lml'm'},a_{lm}|d_{lm})}{\partial \beta^{1M}} = \sum_{l}\frac{\partial A^{1M}_{ll+1}}{\partial \beta^{1M}}\frac{\partial \ln P(S_{lml+1m'},a_{lm}|d_{lm})}{\partial A^{1M}_{ll+1}}
\label{Eq:51}
\end{eqnarray}
and
\begin{equation}
\dot{\beta}^{1M} = \frac{p_{{\beta}^{1M}}}{m_{{\beta}^{1M}}}\,.
\end{equation}
\noindent We take the mass parameter $m_{\beta^{1M}}=1$. We can integrate these equations and infer the posterior of $\beta^{1M}$. In Fig.~(\ref{fig:beta1}) we inferred $\beta^{1M}$
from a SI violated  map generated using CoNIGS~\cite{CoNIGS}. The known Doppler boost injected corresponds to $\beta^{10} = - 1.87\times 10^{-3}$  and 
$\beta^{11} = - 1.24\times 10^{-4} + 1.18\times 10^{-3}i$. We add isotropic Gaussian random noise with $\sigma_n=20\mu K$. 
For our analysis, we remove low multipoles and consider the sum of Eq.(\ref{Eq:51}) from $201$ to $1024$ multipoles in the range, because at the low multipoles, signal is small but the errorbars being large, introduces  unwarranted higher 
error in determining beta. 
Note that the recovery of the Doppler signal is expected to improve as  one includes higher multipoles up to $l\sim 2000$ now available from Planck. Here, in this demonstrative example on a simulated map,  we restrict to $l_{max} =1000$, and also add a noise much higher than that on current CMB experiments to establish the method in a more adverse situation. The analysis demonstrates that models with small number of parameters can be inferred from high-resolution data. 


\section{Discussion and Conclusion}
\label{concl}

We design a general method  to infer the underlying covariance  structure of a random field measured on a sphere
using a completely Bayesian technique.  We employ the Bipolar Spherical Harmonic representation
of the general covariance matrix underlying random fields on a sphere and outline the method for   a fully
Bayesian inference of the angular power spectrum and the  BipoSH coefficients simultaneously from 
a single observed map. We use Hamiltonian Monte Carlo for sampling the posterior distributions of
the BipoSH parametrization of the covariance.

We demonstrate the method with application to simulated CMB sky maps motivated by the
 need for Bayesian assessment of  the presence of the SI violation signal in  observed CMB maps. 
We consider CMB sky maps that are statistically isotropic (SI), as well as, non-SI case studies  that carry signatures of  
a few known cases of SI violation.  We test our method at different noise levels.
Our method recovers  the angular power spectra $C_{l}$ up to high multipoles, $l$,  even
in presence of large noise. The recovery of the BipoSH spectra is also at good fidelity at all the multipoles up to $ l \sim 1000$. 
In case of the non-SI Doppler boost CMB maps, we also carry out a direct Bayesian inference of the posterior distribution of the the governing boost paramter, $\vec{\beta}$. The recovery of the Doppler signal is expected to improve as  one includes higher multipoles up to $l\sim 2000$ now available from Planck and the appropriate noise level that
is much lower than employed here. 
We expect that an application of our method to recently released  exquisite CMB sky maps from Planck will provide reliable 
assessment of SI violation and 
serve as a valuable tool to assess the level of evidence for or agains the candidate anomalies.
Most assessments of non-SI signals have to assume the angular power spectrum. Hence, it is also important  to note that 
here we  jointly infer the angular power spectrum together with BipoSH spectra which makes it specifically
sensitive to non-SI phenomena that affect both, a known example of which is non-trivial topology of the Universe~\cite{Ade2013XXVI,Ade2015XVIII}. 
Besides, hunting for cosmic signal of SI violation, the method  provides an excellent diagnostic of non-SI residuals
originating  from  the handling of unavoidable observational tasks,  such as, removal of foreground emission, noise inhomogeneity and
non symmetric beam response functions. The method should be readily extendable beyond  scalar fields to  CMB polarisation 
maps that are already available on the `full' sky  and also to shear field maps of future weak lensing  observations over large fractions of the sky.

 The general, principled, approach to a Bayesian inference of the covariance structure in a random field on a sphere presented here 
 has broad potential for  application to other many aspects of  cosmology and astronomy, as well as, more distant areas of research 
 like geosciences and climate modelling.
 
\section*{Acknowledgments}
S.D. acknowledge the Council of Scientific and Industrial Research (CSIR), India for
financial support through Senior Research fellowships  and ILP Paris for supporting the collaboration related visit. BDW is supported by a senior Excellence Chair by the Agence Nationale de Recherche (ANR-10-CEXC-004-01) and a Chaire Internationale at the Universit\'e Pierre et Marie Curie. We thank Suvodip Mukherjee for providing nonSI simulations of doppler boost case using CoNIGS. Computations were carried out at the HPC facilities in IUCAA.

 \bibliographystyle{JHEP}
\bibliography{reference}

\end{document}